\documentstyle[12pt]{article}

\begin{document}
 \thispagestyle{empty}
 \begin{flushright}
 {\tt University of Bergen, Department of Physics}    \\[2mm]
 {\tt Scientific/Technical Report No.1995-16}    \\[2mm]
 {\tt ISSN 0803-2696} \\[5mm]
 {hep-ph/9605348} \\[5mm]
\end{flushright}
 \vspace*{3cm}
 \begin{center}
 {\bf \large
 Three-gluon vertex in arbitrary gauge and dimension}
 \end{center}
 \vspace{1cm}
 \begin{center}
 A.~I.~Davydychev$^{a,}$\footnote{davyd@vsfys1.fi.uib.no.
                                    Permanent address:
                                    Institute for Nuclear Physics,
                                    Moscow State University,
                                    119899, Moscow, Russia} , \ \
 P.~Osland$^{a,}$\footnote{Per.Osland@fi.uib.no} \ \
 \ \ and \ \
 O.~V.~Tarasov$^{b,c,}$\footnote{tarasov@ifh.de.
                                 On leave from
                                 Joint Institute for Nuclear Research,
                                 141980, Dubna, Russia.
                                 Present address: IfH, DESY-Zeuthen,
                                 Platanenallee 6, D-15738 Zeuthen, 
                                 Germany.}\\
 \vspace{1cm}
$^{a}${\em
 Department of Physics, University of Bergen, \\
      All\'{e}gaten 55, N-5007 Bergen, Norway}
\\
\vspace{.3cm}
$^{b}${\em
 Fakult\"at f\"ur Physik,
 Universit\"at Bielefeld, \\
 D-33615, Bielefeld 1, Germany}
\\
\vspace{.3cm}
$^{c}${\em
 NORDITA, Blegdamsvej 17, DK-2100 Copenhagen \O, Denmark}
\end{center}
 \hspace{3in}
 \begin{abstract}
{One-loop off-shell contributions to the three-gluon vertex are calculated,
in arbitrary covariant gauge and in arbitrary space-time dimension,
including quark-loop contributions (with massless quarks).
It is shown how one can get the results for all on-shell limits of interest 
directly from the general off-shell expression.
The corresponding general expressions for the one-loop ghost-gluon vertex 
are also obtained. 
They allow for a check of consistency
with the Ward--Slavnov--Taylor identity. }
 \end{abstract}

\newpage

\pagestyle{myheadings}
\section{Introduction}
\setcounter{equation}{0}

The three-gluon coupling is perhaps the most obvious manifestation
of the non-Abelian aspect of Quantum Chromodynamics \cite{QCD}
(see also the reviews \cite{MarPag,CTEQ}).
Implicitly, it has been studied experimentally through the
observed running of the coupling constant \cite{Runningalphas}.
The associated Casimir invariant has even been measured directly
in studies of four-jet events at LEP \cite{Casimir},
the SU(3) group being consistent with the data.

Apart from being a standard object of consideration in
textbooks on Quantum Field Theory and QCD 
(see e.g.\ \cite{IZ,PasTar,Muta}), the perturbative corrections 
to gluonic vertices are also very important in real physical 
calculations, such as multijet production at the hadron colliders
(see e.g.\ \cite{jets,CTEQ} and references therein). At the present
level of accuracy, one needs to perform not only calculations 
with on-shell external particles, 
there are also contributions where general off-shell results
are needed. 

One of the original reasons the three-gluon vertex was studied
was the belief that
its infrared properties might shed light on
the mechanism of confinement.
In these studies, different approaches 
were used, some of which are discussed in
the review \cite{infrared} (and references therein).

For special cases, the one-loop results for the three-gluon
coupling have been known for many years.
Celmaster and Gonsalves (CG) presented in 1979 \cite{CG}
the one-loop result for the vertex, for off-shell gluons,
restricted to the symmetric case, $p_1^2=p_2^2=p_3^3$,
in an arbitrary covariant gauge\footnote{The result of \cite{CG}
was also confirmed by Pascual and Tarrach \cite{PT}.}.
Ball and Chiu (BC) then in 1980 considered the general off-shell case,
but restricted to the Feynman gauge \cite{BC2}.
Later, various on-shell results have also been given,
by Brandt and Frenkel (BF) \cite{BF}, 
restricted to the infrared-singular parts only 
(in an arbitrary covariant gauge), 
and by Nowak, Prasza{\l}owicz and S{\l}omi{\'n}ski (NPS)
\cite{NPS}, who also gave the finite parts for the case of
two gluons being on-shell (in Feynman gauge). 
An overview of these results is given in the table below.

\begin{table}[h]
\begin{center}
\begin{minipage}{148mm}
{\renewcommand{\arraystretch}{1.5}
\begin{tabular}{||l||l|l|l|l||}
\hline
          &\multicolumn{2}{|c|}{\parbox[c]{40mm}{all momenta
                                        off-shell\phantom{${}_I$}}}
          &\multicolumn{2}{|c||}{\parbox[c]{58mm}{some momenta
                                        on-shell\phantom{${}_I$}}} \\
           \cline{2-5}
\phantom{$P^2$} &general case &$p_1^2=p_2^2=p_3^2$
                &$p_3^2=0$ &$p_1^2=p_2^2=0$ \\
\hline \hline
\parbox[c]{20mm}{Feynman \\ gauge}
              &\parbox[c]{22mm}{\raggedright BC \cite{BC2},\phantom{\Large I}
               eq.~(3.3), \\ no quarks}
              &\parbox[c]{22mm}{\raggedright special case of CG \cite{CG}}
              &\parbox[c]{22mm}{\raggedright special case of BF \cite{BF}}
              &\parbox[c]{22mm}{\raggedright NPS~\cite{NPS}, \\ App.~B} \\
\hline
\parbox[c]{25mm}{Arbitrary \\ covariant \\ gauge}
              &
              &\parbox[c]{22mm}{\raggedright CG
                                \cite{CG},\phantom{\Large I} eq.~(14)}
              &\parbox[c]{27mm}{\raggedright BF \cite{BF},\phantom{\Large I}\\
                                eq.~(25),  \\ no quarks, \\
                                no~finite~parts}
              &\parbox[c]{27mm}{\raggedright BF \cite{BF},\phantom{\Large I}\\
                                eq.~(30),  \\ no quarks, \\
                                no~finite~parts} \\
\hline
\end{tabular}
} 
\end{minipage}
\end{center}
\caption{Kinematics and gauges considered in other studies}
\end{table}

From this table one can see that, even if we consider the 
results in (or around) four dimensions, there are still several
``white spots''. They correspond not only to the most general case 
(the lower left corner), but also to some other cases when
the results are missing, either for quark loop contributions 
or for the finite parts. 
The aim of the present paper is to cover {\em all} such remaining spots
(for the case when massless quarks are considered).
Moreover, we present results which are valid 
for an {\em arbitrary} value of the space-time dimension.
Apart from the three-gluon vertex itself, we also consider the
ghost-gluon vertex and two-point functions, to be able to check
that all these expressions obey the Ward--Slavnov--Taylor
identity for the three-gluon vertex.

At the one-loop level, the simple and well-known Lorentz structure
of the lowest-order coupling gets modified. In the general case, six 
tensor structures (and their permutations) are needed to decompose
the three-gluon vertex \cite{BC2}. 
Thus, six scalar functions multiplying these tensor structures
are to be calculated.
These scalar functions depend on
the gauge parameter,
the space-time dimension, and the kinematical invariants
($p_1^2$, $p_2^2$, $p_3^2$).

There are several reasons why the one-loop results calculated in
arbitrary gauge and dimension $n$ are of special interest: \\
(i) knowing the results in arbitrary gauge, one can explicitly
keep track of gauge invariance for physical quantities; \\
(ii) if one is interested in the two-loop calculation of the three-gluon 
coupling, one should know one-loop contributions in more detail; \\
(iii) results in arbitrary dimension make it possible to consider
all on-shell limits (when some $p_i^2=0$) {\em directly} from
these expressions (see section~4), this is impossible if one only has
the results valid around four dimensions; \\
(iv) QCD is also a theory of interest in three and two dimensions 
(see e.g.\ \cite{Teper} and the review \cite{QCD2}); \\
(v) as we shall see, the results for arbitrary dimension are not much 
more cumbersome than those considered around four dimensions (in some
respects, they are even more transparent and instructive).  
  
We note that in several papers the one-loop three-gluon
vertex in axial-type gauges (including the light-cone gauge) was considered
\cite{axial} (mainly divergent parts and special limits have been
studied). The three-gluon vertex in the background field formalism
was considered in ref.~\cite{Friedman}, while the gauge-invariant
vertex was studied in ref.~\cite{Cornwall}. 
Moreover, there were some lattice calculations  of the three-gluon 
vertex, see e.g.\ ref.~\cite{lattice}.
We shall not address
these issues here, but concentrate instead on the {\em standard}
vertex in an arbitrary {\em covariant} gauge. 

The paper is organized as follows. 
In section~2, we introduce the notation for the two- and three-point
functions to be considered, and discuss their decomposition in terms of 
scalar functions as well as the corresponding 
Ward--Slavnov--Taylor identity.
In section~3, we present the most general off-shell results
for the three-gluon vertex. Section~4 contains the corresponding 
expressions for all on-shell limits of interest. In section~5, 
we conclude with
a summary and a discussion of the results.
Then, we have several appendices where some further results
and technical details are presented, such as the formulae used
to decompose the three-gluon vertex (Appendix~A), relevant results for
the scalar integrals involved (Appendix~B), results for
the self energies (Appendix~C) and the ghost-gluon vertex (Appendix~D),
expressions for the on-shell limit $p_3^2=0$ in arbitrary gauge
(Appendix~E), and also some results for $p_1^2=p_2^2=0$ (Appendix~F).      

\section{Preliminaries}
\setcounter{equation}{0}

The Yang--Mills term of the QCD Lagrangian yields the 
following well-known expression for the lowest-order three-gluon vertex:
\begin{equation}
\label{2eq:tree-level}
-\mbox{i} \; g \; f^{a_1 a_2 a_3} 
\left[ g_{\mu_1 \mu_2} (p_1 - p_2)_{\mu_3}
      + g_{\mu_2 \mu_3} (p_2 - p_3)_{\mu_1}
      + g_{\mu_3 \mu_1} (p_3 - p_1)_{\mu_2}
\right] , 
\end{equation}
where $p_1, p_2$ and $p_3$ are the momenta of the gluons,
    all of which are      ingoing, $p_1+p_2+p_3=0$.
In (\ref{2eq:tree-level}), the $f^{a_1 a_2 a_3}$ 
are the totally antisymmetric colour structures corresponding
to the adjoint representation of the gauge group\footnote{Although
the standard QCD Lagrangian corresponds to the $\mbox{SU}(3)$
group, our results are valid for an arbitrary semi-simple
gauge group.}.
They can
be extracted from the general three-gluon vertex by 
defining\footnote{In fact, also completely symmetric
colour structures $d^{a_1 a_2 a_3}$ might be considered,
but they do not appear in the perturbative calculation of
QCD three-point vertices at the one-loop level.}  
\begin{equation}
\label{ggg}
\Gamma_{\mu_1 \mu_2 \mu_3}^{a_1 a_2 a_3}(p_1, p_2, p_3)
\equiv  - \mbox{i} \; g \;
f^{a_1 a_2 a_3} \; \Gamma_{\mu_1 \mu_2 \mu_3}(p_1, p_2, p_3) .
\end{equation}
Since the gluons are bosons, and since the colour structures
$f^{a_1 a_2 a_3}$ are antisymmetric, 
$\Gamma_{\mu_1 \mu_2 \mu_3}(p_1, p_2, p_3)$ must also be {\em antisymmetric}
under any interchange of a pair of gluon momenta and the corresponding
Lorentz indices.

The lowest-order gluon propagator is
\begin{equation}
\label{gl_prop}
\delta^{a_1 a_2} \; \frac{1}{p^2} 
\left( g_{\mu_1 \mu_2} - \xi \; \frac{p_{\mu_1} p_{\mu_2}}{p^2} 
\right) ,
\end{equation}
where $\xi$ is the gauge parameter corresponding to a general
covariant gauge, defined such that $\xi=0$ is the Feynman gauge.
Here and henceforth, a causal prescription is understood,
$1/p^2 \rightarrow 1/(p^2 +\mbox{i}0)$.
 
When one calculates radiative corrections to the three-gluon vertex
(the corresponding one-loop diagrams are presented in fig.~1), 
other tensor structures
arise, in addition to the lowest-order expression 
(\ref{2eq:tree-level}), and the general tensor decomposition 
should be considered. If we take into account momentum
conservation (only two of the external momenta are independent), 
fourteen independent tensor structures carrying three Lorentz indices exist,
and in general $\Gamma_{\mu_1 \mu_2 \mu_3}$ can be written as a sum of
these tensors multiplied by scalar functions (see Appendix A).
This decomposition is useful for extracting the corresponding scalar
functions from the result of a calculation.
Although bosonic symmetry of the vertex puts some conditions
on the corresponding scalar functions, the explicit symmetry of
the expression is broken, because one of the momenta was substituted
in terms of two others.

To avoid this, one can use a more symmetric decomposition of the general 
three-gluon vertex, proposed by Ball and Chiu \cite{BC2}\footnote{Another 
general decomposition of the three-gluon vertex was considered 
in ref.~\cite{KB}.},
\begin{eqnarray}
\label{BC-ggg}
\Gamma_{\mu_1 \mu_2 \mu_3}(p_1, p_2, p_3)
= A(p_1^2, p_2^2; p_3^2)\; g_{\mu_1 \mu_2} (p_1-p_2)_{\mu_3}
+ B(p_1^2, p_2^2; p_3^2)\; g_{\mu_1 \mu_2} (p_1+p_2)_{\mu_3}
\hspace{10mm}
\nonumber \\
- C(p_1^2, p_2^2; p_3^2)\;
\left( (p_1 p_2) g_{\mu_1 \mu_2} - {p_1}_{\mu_2} {p_2}_{\mu_1} \right)
\; (p_1-p_2)_{\mu_3}
\hspace{55mm}
\nonumber \\
+ \textstyle{1\over3}\; S(p_1^2, p_2^2, p_3^2)
\; \left( {p_1}_{\mu_3} {p_2}_{\mu_1} {p_3}_{\mu_2}
        + {p_1}_{\mu_2} {p_2}_{\mu_3} {p_3}_{\mu_1} \right)
\hspace{62mm}
\nonumber \\
+ F(p_1^2, p_2^2; p_3^2)\;
\left( (p_1 p_2) g_{\mu_1 \mu_2} - {p_1}_{\mu_2} {p_2}_{\mu_1} \right)
\; \left( {p_1}_{\mu_3} (p_2 p_3) - {p_2}_{\mu_3} (p_1 p_3) \right)
\hspace{28mm}
\nonumber \\
+ H(p_1^2, p_2^2, p_3^2)
\left[ -g_{\mu_1 \mu_2}
\left( {p_1}_{\mu_3} (p_2 p_3) \!-\! {p_2}_{\mu_3} (p_1 p_3) \right)
+ \textstyle{1\over3}
\left( {p_1}_{\mu_3} {p_2}_{\mu_1} {p_3}_{\mu_2}
\!-\! {p_1}_{\mu_2} {p_2}_{\mu_3} {p_3}_{\mu_1} \right) \right]
\nonumber \\
+ \left\{ \; \mbox{cyclic permutations of} \; (p_1,\mu_1),
(p_2,\mu_2), (p_3,\mu_3)\; \right\}  .
\hspace{20mm}
\end{eqnarray}
Here, the $A$, $C$ and $F$ functions are symmetric in the first two
arguments, the $H$ function is totally symmetric, the $B$ function
is antisymmetric
in the first two arguments, while the $S$ function is antisymmetric
with respect to interchange of any pair of arguments.
Note that the contribution containing the $F$ and $H$ functions
is totally transverse, i.e. it gives zero when contracted with
any of ${p_1}_{\mu_1}$, ${p_2}_{\mu_2}$ or ${p_3}_{\mu_3}$.

Now, before proceeding further, we introduce some notation.
For a quantity $X$ (e.g. any of the scalar functions contributing to the
propagators or the vertices), we shall denote the zero-loop-order 
contribution as $X^{(0)}$, and the one-loop-order contribution as
$X^{(1)}$. In this paper, as a rule, 
\begin{equation}
X^{(1)}=X^{(1,\xi)} + X^{(1,q)},
\end{equation}
where $X^{(1,\xi)}$ denotes a contribution of gluon and ghost 
loops in a general covariant gauge (\ref{gl_prop}) (in particular, 
$X^{(1,0)}$ corresponds to the Feynman gauge, $\xi=0$),
while $X^{(1,q)}$ represents the contribution of the
quark loops. 

For example, from (\ref{2eq:tree-level}) one can see that at 
the ``zero-loop'' level
all the scalar functions involved in (\ref{BC-ggg}) vanish,
except the $A$ function which is
\begin{equation}
\label{A0=1}
A^{(0)} = 1 .
\end{equation}

In what follows, we shall also need to use some other QCD Green functions,
including those involving the Faddeev--Popov ghosts.
As in (\ref{BC-ggg}) we define the corresponding scalar structures
following the notation of the paper \cite{BC2}. 

The gluon polarization operator is defined as
\begin{equation}
\label{gl_po}
\Pi_{\mu_1 \mu_2}^{a_1 a_2}(p) 
\equiv - \delta^{a_1 a_2} 
\left( p^2 g_{\mu_1 \mu_2} - {p}_{\mu_1}{p}_{\mu_2} \right) J(p^2),
\end{equation}
while the ghost self energy is
\begin{equation}
\label{gh_se}
\widetilde{\Pi}^{a_1 a_2}(p^2) = \delta^{a_1 a_2} \; p^2 \; G(p^2) .
\end{equation}
The lowest-order results are $J^{(0)}=G^{(0)}=1$.
The one-loop contributions to $\Pi_{\mu_1 \mu_2}^{a_1 a_2}(p)$ 
and $\widetilde{\Pi}^{a_1 a_2}(p^2)$ are presented 
in fig.~2 and can easily be calculated.
The results (in arbitrary space-time dimension) can be found
e.g.\ in ref.~\cite{Muta}. For completeness, we collect the relevant
formulae in Appendix~C.

The ghost-gluon vertex can be represented as
\begin{equation}
\label{ghg}
\widetilde{\Gamma}_{\mu_3}^{a_1 a_2 a_3}(p_1, p_2; p_3)
\equiv -\mbox{i} g \; f^{a_1 a_2 a_3} \;
{p_1}^{\mu} \; \widetilde{\Gamma}_{\mu \mu_3}(p_1, p_2; p_3) ,
\end{equation}
where $p_1$ is the out-ghost momentum, $p_2$ is the in-ghost momentum,
$p_3$ and $\mu_3$ are the momentum and the Lorentz index of the gluon
(all momenta are ingoing).
For $\widetilde{\Gamma}_{\mu \mu_3}$ we adopt the following decomposition,
also used in \cite{BC2}:
\begin{eqnarray}
\label{BC-ghg}
\widetilde{\Gamma}_{\mu \mu_3}(p_1,p_2;p_3)
= g_{\mu \mu_3} a(p_3,p_2,p_1)
- {p_3}_{\mu} {p_2}_{\mu_3} b(p_3,p_2,p_1)
+ {p_1}_{\mu} {p_3}_{\mu_3} c(p_3,p_2,p_1)
\nonumber \\
+ {p_3}_{\mu} {p_1}_{\mu_3} d(p_3,p_2,p_1)
+ {p_1}_{\mu} {p_1}_{\mu_3} e(p_3,p_2,p_1) .
\end{eqnarray}
At the ``zero-loop'' level,
\begin{equation}
\label{ghg0}
\widetilde{\Gamma}^{(0)}_{\mu \mu_3} = g_{\mu \mu_3} ,
\end{equation}
and therefore all the scalar functions involved in  (\ref{BC-ghg})
vanish at this order, except one, $a^{(0)}=1$.
We shall also need the one-loop-order results for the ghost-gluon
vertex in arbitrary gauge (the corresponding diagrams are presented
in fig.~3). 
We have calculated one-loop contributions to all scalar functions 
occurring on the r.h.s.\ of eq.~(\ref{BC-ghg}), they are presented
in Appendix D.  

We need $\widetilde{\Gamma}_{\mu \mu_3}$ with {\em two} Lorentz 
indices, because
this is what enters the Ward--Slavnov--Taylor identity for the three-gluon
vertex, which,
in the covariant gauge, has the following form (see e.g.
in \cite{MarPag,BC2}):
\begin{eqnarray}
\label{WST}
p_3^{\mu_3} \; \Gamma_{\mu_1 \mu_2 \mu_3}(p_1, p_2, p_3)
= - J(p_1^2) \; G(p_3^2) \;
\left( g_{\mu_1 \;\;\;}^{\;\;\; \mu_3}\; p_1^2
       - {p_1}_{\mu_1} \; {p_1}^{\mu_3} \right) \;
\widetilde{\Gamma}_{\mu_3 \mu_2}(p_1, p_3; p_2)
\hspace{1mm}
\nonumber \\
 + J(p_2^2) \; G(p_3^2) \;
\left( g_{\mu_2 \;\;\;}^{\;\;\; \mu_3}\; p_2^2
       - {p_2}_{\mu_2} \; {p_2}^{\mu_3} \right) \;
\widetilde{\Gamma}_{\mu_3 \mu_1}(p_2, p_3; p_1) .
\end{eqnarray}
It is easy to see that the $F$ and $H$ functions from the three-gluon vertex
(\ref{BC-ggg}), as well as the $c$ and $e$ functions
from the ghost-gluon vertex (\ref{BC-ghg}) do not contribute
to this identity.
Below, we are going to use eq.~(\ref{WST}) as a non-trivial check on
the results for the longitudinal part of the three-gluon vertex. 

To conclude this section, we would like to present the notation
we use for the integrals occurring in the one-loop calculations.
We define the integral corresponding to the 
triangle diagram as
\begin{equation}
\label{defJ}
J (\nu_1  ,\nu_2  ,\nu_3) \equiv \int
 \frac{\mbox{d}^n q}{ ((p_2 -q )^2)^{\nu_1}  ((p_1 +q )^2)^{\nu_2}
      (q^2)^{\nu_3} } ,
\end{equation}
where $n=4-2\varepsilon$ is the space-time dimension (in the framework
of dimensional regulariza\-tion\footnote{For simplicity, we put the 
dimensional-regularization scale $\mu_{DR} = 1$. Otherwise,
all one-loop expressions for dimensionally-regularized quantities
should have been multiplied by $(\mu_{DR})^{2\varepsilon}$.
In the final results, expanded around $n=4$ and renormalized,
this scale can easily be restored by inserting $\mu_{DR}$
in all non-dimensionless arguments of the logarithms,
in order to make them dimensionless. See also section~3.5
where the renormalization is discussed.}, 
\cite{dimreg}).  
A brief overview of relevant results for such integrals in $n$
dimensions is presented in Appendix B. 
It should be noted that all such integrals occurring in the
present calculation can be algebraically reduced to 
one non-trivial integral,
\begin{equation}
\label{J(1,1,1)}
J(1,1,1) = {\mbox{i}} \pi^{n/2} \;
\eta \;
\varphi(p_1^2,p_2^2,p_3^2) ,
\end{equation}
where $\varphi(p_1^2,p_2^2,p_3^2) \equiv \varphi$ is a totally
symmetric function (see Appendix B for details),
and three two-point integrals, $J(0,1,1)$, $J(1,0,1)$ and $J(1,1,0)$, 
which can be expressed in terms of a power-like function
\begin{equation}
\label{kappa}
\kappa(p_i^2) \equiv \kappa_i
= - \frac{2}{(n-3) (n-4)} \; (-p_i^2)^{(n-4)/2}
= \frac{1}{\varepsilon (1-2\varepsilon)} \; (-p_i^2)^{-\varepsilon}
\end{equation}
as, e.g.,
\begin{equation}
\label{J(1,1,0)}
J(1,1,0) = {\mbox{i}} \pi^{n/2} \;
\eta \;
\kappa(p_3^2) ,
\end{equation}
and similarly for $J(0,1,1)$ and $J(1,0,1)$, where, instead of
$\kappa(p_3^2)=\kappa_3$, we should use 
$\kappa(p_1^2)=\kappa_1$ and $\kappa(p_2^2)=\kappa_2$,
respectively. In eqs.~(\ref{J(1,1,1)}) and (\ref{J(1,1,0)}),
$\eta$ denotes a factor constructed of $\Gamma$ functions,
\begin{equation}
\label{eta}
\eta \equiv 
\frac{\Gamma^2(\frac{n}{2}-1)}{\Gamma(n-3)} \; 
     \Gamma(3-{\textstyle{n\over2}}) =
\frac{\Gamma^2(1-\varepsilon)}{\Gamma(1-2\varepsilon)} \; 
\Gamma(1+\varepsilon) .
\end{equation}


\section{Off-shell results}
\setcounter{equation}{0}

The set of Feynman diagrams yielding one-loop contributions
to the three-gluon vertex is presented in Fig.~1.

When calculating the diagrams, we used the standard technique
of tensor decomposition \cite{tensor}\footnote{An alternative way
to decompose triangle integrals (\ref{defJ}) with tensor numerators
was used in \cite{HS}.
It was based on a formula from \cite{PLB'91}.}, reducing the result to
combinations of scalar integrals multiplying the tensor
structures constructed from the external momenta (see Appendix~A).
In the Feynman gauge, the basic set of scalar integrals (\ref{defJ})
includes the four integrals $J(1,1,1)$, $J(0,1,1)$, $J(1,0,1)$
and $J(1,1,0)$ only, since massless integrals with two non-positive
powers $\nu_i$ vanish in dimensional regularization \cite{dimreg}. 
For arbitrary $\xi$, we also get integrals with some of the powers
of the denominators equal to two, see eq.~(\ref{gl_prop}). 
However, with the help of the integration-by-parts technique \cite{ibp}
these integrals can be algebraically reduced to the above basic set
(see ref.~\cite{JPA}). While performing the calculations,
the {\sf REDUCE} system \cite{reduce} was heavily employed.

Before presenting the results, let us define two totally symmetric
combinations of the invariants formed from the external momenta,
\begin{equation}
\label{qqq}
{\cal{Q}} \equiv (p_1 p_2) + (p_1 p_3) + (p_2 p_3)
          = - \textstyle{1\over2} (p_1^2 + p_2^2 + p_3^2) ,
\end{equation}
\begin{eqnarray}
\label{kkk}
{\cal{K}} & \equiv & p_1^2 p_2^2 - (p_1 p_2)^2
                   = p_1^2 p_3^2 - (p_1 p_3)^2
                   = p_2^2 p_3^2 - (p_2 p_3)^2
\nonumber \\
&=& (p_1 p_2)(p_1 p_3) + (p_1 p_2)(p_2 p_3) + (p_1 p_3)(p_2 p_3)
\nonumber \\
&=& - \textstyle{1\over4} \; \left(
(p_1^2)^2 + (p_2^2)^2 + (p_3^2)^2
- 2 p_1^2 p_2^2 - 2 p_1^2 p_3^2 - 2 p_2^2 p_3^2
\right) .
\end{eqnarray}
From the last line of eq.~(\ref{kkk}), one can recognize 
the structure $-4{\cal{K}}$ as  
the K\"allen function of $p_1^2, p_2^2$ and $p_3^2$, 
see e.g. in \cite{Kallen}.

\subsection{Results in the Feynman gauge}

Let us consider the one-loop contributions 
to the three-gluon vertex
(\ref{BC-ggg}) in the Feynman gauge ($\xi=0$), without the quark loops
(the results for the latter are presented in Section~3.3).
We shall use the standard notation $C_A$ for the Casimir constant,
\begin{equation}
\label{C_A}
f^{acd}f^{bcd} = C_A \, \delta^{ab} \hspace{5mm} 
(C_A = N \; \mbox{for the SU($N$) group}),
\end{equation}
whereas the factor $\eta$ occurring in the results is defined by
eq.~(\ref{eta}).
 
The one-loop results for the scalar functions
(\ref{BC-ggg}), for arbitrary value of the
space-time dimension $n$, are  
\begin{eqnarray}
\label{A(1,0)}
A^{(1,0)}(p_1^2, p_2^2; p_3^2) =
\frac{g^2 \; \eta}{(4\pi)^{n/2}} \; C_A \; \frac{1}{4 (n-1) \cal{K}}
\hspace{79mm}
\nonumber \\
\times \left\{ (n-1) \left( p_3^2 + 3 (p_1 p_2)\right)
\left[ p_3^2 (p_1 p_2) \varphi + (p_1 p_3) \kappa_1
      + (p_2 p_3) \kappa_2 + p_3^2 \kappa_3 \right]
\right.
\nonumber \\
\left.
+ 4 (n-1) {\cal{K}} \left[ (p_1 p_2) \varphi + \kappa_3 \right]
- (3n-2) {\cal{K}} \left[ \kappa_1 + \kappa_2 \right]
\right\} ,
\hspace{20mm}
\end{eqnarray}
\begin{eqnarray}
\label{B(1,0)}
B^{(1,0)}(p_1^2, p_2^2; p_3^2) =
- \frac{g^2 \; \eta}{(4\pi)^{n/2}} \; C_A \; 
 \frac{1}{4 (n-1) \cal{K}} (p_1^2 - p_2^2)
\hspace{59mm}
\nonumber \\
\times\!
\left\{ (n\!-\!1)
\left[ (p_1 p_3)(p_2 p_3) \varphi + (p_1 p_3) \kappa_1 
       + (p_2 p_3) \kappa_2 + p_3^2 \kappa_3 \right]
+ (4n\!-\!3) {\cal{K}} \; \frac{\kappa_1-\kappa_2}{p_1^2 \!-\! p_2^2}
\right\} ,
\hspace{3mm}
\end{eqnarray}
\begin{eqnarray}
\label{C(1,0)}
C^{(1,0)}(p_1^2, p_2^2; p_3^2) =
\frac{g^2 \; \eta}{(4\pi)^{n/2}} \; C_A \;
 \frac{1}{4 (n-1) \cal{K}} 
\hspace{77mm}
\nonumber \\
\times
\left\{ 3 (n\!-\!1)
\left[ p_3^2 (p_1 p_2) \varphi + (p_1 p_3) \kappa_1
      + (p_2 p_3) \kappa_2 + p_3^2 \kappa_3 \right]
- 2\; (4n\!-\!3)\; {\cal{K}} \; \frac{\kappa_1-\kappa_2}{p_1^2 \!-\! p_2^2}
\right\} ,
\hspace{3mm}
\end{eqnarray}
\begin{equation}
\label{S(1,0)}
S^{(1,0)}(p_1^2, p_2^2, p_3^2) = 0 ,
\end{equation}
\begin{eqnarray}
\label{F(1,0)}
F^{(1,0)}(p_1^2, p_2^2; p_3^2) = 
\frac{g^2 \; \eta}{(4\pi)^{n/2}} \; C_A \;
\frac{1}{4 (n-1) {\cal{K}}^3}
\hspace{77mm}
\nonumber \\
\times \! \left\{ \frac{}{}
2  \left[(n^2-1) (p_1 p_2) (p_1 p_3) (p_2 p_3)
+2 (n-2) p_3^2 {\cal{K}}-(n-7) (p_1 p_2) {\cal{K}}\right]
\right.
\hspace{33mm}
\nonumber \\
\times \left[ p_3^2 (p_1 p_2) \varphi + (p_1 p_3) \kappa_1
             +(p_2 p_3) \kappa_2 + p_3^2 \kappa_3 \right]
\hspace{5mm}
\nonumber \\
       +2 {\cal{K}} \left[(n+1) (n-4) (p_1 p_3) (p_2 p_3)
                          -(5 n-11) {\cal{K}}\right] \;
\left[(p_1 p_2) \varphi+\kappa_3\right]
\hspace{31mm}
\nonumber \\
       +2 p_3^2 {\cal{K}} \left[(n+1) (p_1 p_3) (p_2 p_3)
                 +(n-3) {\cal{K}}\right] \varphi
       +(4 n-7) {\cal{K}}^2  \left[ \kappa_1+\kappa_2 \right]
\hspace{27mm}
\nonumber \\
\left. +  {\cal{K}} \left[
       2 (n+1) (p_1 p_2) (p_1^2 - p_2^2)^2
       + (4 n-3)  {\cal{K}}
        (p_1^2 \!+\! p_2^2 \!-\! 2 (p_1 p_2))
\right] \;
\frac{\kappa_1 - \kappa_2}{p_1^2 - p_2^2} 
\right\} ,
\hspace{15mm}
\end{eqnarray}
\begin{eqnarray}
\label{H(1,0)}
H^{(1,0)}(p_1^2, p_2^2, p_3^2) =
\frac{g^2 \; \eta}{(4\pi)^{n/2}} \; C_A \;
\frac{1}{2 (n-1) {\cal{K}}^3}
\hspace{77mm}
\nonumber \\
\times
\left\{ (n^2-1) (p_1 p_2)(p_1 p_3)(p_2 p_3)
\hspace{90mm}
\right.
\nonumber \\
\times
\left[ (p_1 p_2)(p_1 p_3)(p_2 p_3) \varphi + (p_1 p_2)(p_1 p_3) \kappa_1
+ (p_1 p_2)(p_2 p_3) \kappa_2 + (p_1 p_3)(p_2 p_3) \kappa_3 \right]
\nonumber \\
- 3 (n-1) (p_1 p_2)(p_1 p_3)(p_2 p_3) {\cal{K}}
\left[
          {\cal{Q}} \varphi
        +  \kappa_1  +  \kappa_2 +  \kappa_3 \right]
+ 2 (n-1) {\cal{K}}^3 \; \varphi
\hspace{20mm}
\nonumber \\
+ (n-2) {\cal{K}} \left[
p_1^2 \left( p_1^2 (p_2 p_3) + (p_1 p_2)(p_1 p_3) \right) \kappa_1
+ p_2^2 \left( p_2^2 (p_1 p_3) + (p_1 p_2)(p_2 p_3) \right) \kappa_2
\right.
\nonumber \\
\left. \left.
+ p_3^2 \left( p_3^2 (p_1 p_2) + (p_1 p_3)(p_2 p_3) \right) \kappa_3
\right] \right\} .
\hspace{10mm}
\end{eqnarray}

When expanded around $n=4$, these formulae
coincide\footnote{Up to the definition of the renormalization
scheme constant $C$ in \cite{BC2}, which we find to be
$C~=~-~\gamma~-~\ln\pi~+~2$ rather than $C=-\gamma-\ln\pi$.}
with the results presented in \cite{BC2}. 
We shall see
that the result $S=0$ is valid also in an arbitrary gauge.
It should be noted that presenting the results in arbitrary
dimension does not spoil their compactness, as compared with
the formulae expanded around $n=4$.

\subsection{Results in arbitrary covariant gauge}

In an arbitrary gauge, the results for the scalar functions of the 
three-gluon vertex (\ref{BC-ggg}) are obviously less compact
than those in the Feynman gauge. We list them below,
also for arbitrary value of the space-time dimension: 
\begin{eqnarray}
\label{A(1,xi)}
A^{(1, \xi)}(p_1^2, p_2^2; p_3^2)
= \frac{g^2 \; \eta}{(4\pi)^{n/2}} \; C_A \; 
\frac{1}{32 {\cal{K}}^2 \; p_1^2 \; p_2^2}
\hspace{78mm}
\nonumber \\
   \times\!\left\{
\left[ p_1^2 p_2^2 {\cal{K}} \left(
        \left( 8-4 \xi-(n-2) (n-3) \xi^2 \right) p_3^2
       +2 \left( 12+4 (n-3) \xi+(n-3) \xi^2 \right) (p_1 p_2) \right)
\right. \right.
\hspace{2mm}
\nonumber \\
\left.
\!+\! \xi \left( (n\!-\!4) \xi\!+\!4 \right) {\cal{K}} {\cal{Q}}
\left( (n\!-\!3) (p_1 p_2) {\cal{Q}}-(n\!-\!4) {\cal{K}} \right)
\!+\! \xi \left( (n\!-\!3) \xi\!+\!2 \right)
(n\!-\!1) p_1^2 p_2^2 p_3^2 (p_1 p_2) {\cal{Q}}
\right]
\hspace*{-3mm}
\nonumber \\
\times \left[ p_3^2 (p_1 p_2) \varphi +(p_1 p_3) \kappa_1 
           +(p_2 p_3) \kappa_2 +p_3^2 \kappa_3 \right]
\nonumber \\
 -{\cal{K}} 
\left[ \left( (n-4) \xi+4 \right) {\cal{K}}
\left( \left( (n-4) \xi-8 \right) p_1^2 p_2^2
      +\xi {\cal{Q}} \left( (n\!-\!2) p_3^2-2 (n\!-\!3) (p_1 p_2) \right)
\right)
\right.
\hspace{14mm}
\nonumber \\
\left.
-\xi \left( (n\!-\!3) \xi+2 \right) (n\!-\!2) p_1^2 p_2^2 p_3^2 {\cal{Q}}
\right]
\left[ (p_1 p_2) \varphi+\kappa_3 \right]
\nonumber \\
+{\cal{K}}\; \varphi 
\left[ \xi \left( (n-4) \xi+4 \right)
{\cal{K}} \left( (2 n-7) p_1^2 p_2^2 p_3^2
                +{\cal{Q}} \left( p_1^2 (p_1 p_3)+p_2^2 (p_2 p_3) \right)
                           \right)
\right.
\hspace{25mm}
\nonumber \\
\left.
+\xi \left( (n-3) \xi+2 \right) p_1^2 p_2^2 p_3^2
   \left( p_3^2 {\cal{Q}}-2 (n-4) {\cal{K}} \right) \right] 
\nonumber \\
-\frac{{\cal{K}}}{n-1} \left[ p_1^2 p_2^2 {\cal{K}}
 \left( 8 (3 n-2)+4 (n-1) (5 n-17) \xi-3 (n-1) (n-4) \xi^2 \right)
\right.
\hspace{24mm}
\nonumber \\
\left.
\left.
+\xi \left( (n-4) \xi+4 \right) (n-1) {\cal{K}} {\cal{Q}}^2
+\xi ((n-3) \xi+2) (n-1) p_1^2 p_2^2 p_3^2 {\cal{Q}} \right]
\left[ \kappa_1+\kappa_2 \right] \right\} ,
\hspace{3mm}
\end{eqnarray}
\begin{eqnarray}
\label{B(1,xi)}
B^{(1, \xi)}(p_1^2, p_2^2; p_3^2)
= \frac{g^2 \; \eta}{(4\pi)^{n/2}} \; C_A \; 
\frac{1}{32 {\cal{K}}^2 \; p_1^2 \; p_2^2 \; p_3^2} \; (p_1^2-p_2^2)
\hspace{55mm}
\nonumber \\
\times\! \left\{ \left[
-p_1^2 p_2^2 p_3^2 {\cal{K}} \left( 8-12 \xi-(n+2) (n-3) \xi^2 \right)
\right. \right.
\hspace{73mm}
\nonumber \\
+\xi \left( (n-4) \xi+4 \right) {\cal{K}} {\cal{Q}}
\left( (n-2) {\cal{K}}+(n-3) (p_1 p_2) (p_3^2+(p_1 p_2)) \right)
\hspace{35mm}
\nonumber \\
\left.
+\xi \left( (n-3) \xi+2 \right) p_1^2 p_2^2
 \left( 2 {\cal{K}} {\cal{Q}} + (n-1) p_3^2 (p_1 p_2) (p_3^2+(p_1 p_2))
\right) \right]
\hspace{36mm}
\nonumber \\
\times \left[ p_3^2 (p_1 p_2) \varphi+ (p_1 p_3) \kappa_1 
             +(p_2 p_3) \kappa_2+ p_3^2 \kappa_3 \right]
\nonumber \\
+(n-4) {\cal{K}} \left[ \xi \left( (n-4) \xi+4 \right)
{\cal{K}} \left( p_3^2 {\cal{Q}}+p_1^2 p_2^2 \right)
+ \xi \left( (n-3) \xi+2 \right) p_1^2 p_2^2 p_3^2 (p_3^2+(p_1 p_2))
\right]
\nonumber \\
\times \left[ (p_1 p_2) \varphi+\kappa_3 \right]
\nonumber \\
-{\cal{K}} p_3^2 \varphi \left[ p_1^2 p_2^2 {\cal{K}}
\left( 8+4 (n-5) \xi-(3 n-10) \xi^2 \right)
\right.
\hspace{63mm}
\nonumber \\
\left.
+\xi \left( (n-4) \xi+4 \right) {\cal{K}} {\cal{Q}}^2
-\xi \left( (n-3) \xi+2 \right) p_1^2 p_2^2 p_3^2 (p_3^2+(p_1 p_2))
\right] 
\nonumber \\
- \frac{{\cal{K}}}{n-1} \; \frac{\kappa_1-\kappa_2}{p_1^2\!-\!p_2^2}
 \left[ p_1^2 p_2^2 p_3^2 {\cal{K}}
       \left( 8 (4 n-3)+4 (n-1) (5 n-19) \xi-(n-1) (5 n-18) \xi^2 \right)
\right.
\nonumber \\
+\xi \left( (n-4) \xi+4 \right)
(n-1) {\cal{K}} {\cal{Q}} \left( p_1^2 (p_1 p_3)+p_2^2 (p_2 p_3) \right)
\hspace{25mm}
\nonumber \\
\left.
\left.
-\xi \left( (n-3) \xi+2 \right) (n-1) p_1^2 p_2^2
   (p_1^2-p_2^2)^2 ( p_3^2+(p_1 p_2)) \right]
\right\} ,
\hspace{3mm}
\end{eqnarray}
\begin{eqnarray}
\label{C(1,xi)}
C^{(1, \xi)}(p_1^2, p_2^2; p_3^2)
= \frac{g^2 \; \eta}{(4\pi)^{n/2}}\; C_A \; 
\frac{1}{16 {\cal{K}}^2 \; p_1^2 \; p_2^2 \; p_3^2}
\hspace{72mm}
\nonumber \\
\times\! \left\{ \frac{}{} \left[
2 p_1^2 p_2^2 p_3^2 {\cal{K}} \left( 6+(2 n-5) \xi+(n-3) \xi^2 \right)
\right. \right.
\hspace{74mm}
\nonumber \\
\left.
+\xi \left( (n\!-\!4) \xi+4 \right) {\cal{Q}} {\cal{K}}
\left( {\cal{K}}+(n\!-\!3) p_3^2 (p_1 p_2) \right)
+\xi \left( (n\!-\!3) \xi+2 \right) (n\!-\!1) p_1^2 p_2^2 (p_3^2)^2 (p_1 p_2)
\right]
\nonumber \\
\times \left[ p_3^2 (p_1 p_2) \varphi+(p_1 p_3) \kappa_1 
             +(p_2 p_3) \kappa_2+p_3^2 \kappa_3 \right]
\nonumber \\
+(n-4) {\cal{K}} p_3^2
\left[ \xi \left( (n\!-\!4) \xi \!+\! 4 \right) {\cal{K}} {\cal{Q}}
      +\xi \left( (n\!-\!3) \xi \!+\! 2 \right) p_1^2 p_2^2 p_3^2 \right]
  \left[ (p_1 p_2) \varphi+\kappa_3 \right]
\hspace{20mm}
\nonumber \\
+{\cal{K}} p_3^2 \; \varphi \left[
\xi \left( (n-4) \xi+4 \right) {\cal{K}}
    \left( p_1^2 p_2^2-{\cal{Q}}^2 \right)
+\xi \left( (n-3) \xi+2 \right) p_1^2 p_2^2 (p_3^2)^2 \right] 
\hspace{24mm}
\nonumber \\
-\frac{{\cal{K}}}{n-1} \; \frac{\kappa_1-\kappa_2}{p_1^2\!-\!p_2^2}
  \left[ 2 p_1^2 p_2^2 p_3^2 {\cal{K}}
  \left(4 (4 n\!-\!3)+2 (n\!-\!1) (5 n\!-\!18) \xi
-(n\!-\!1) (2 n\!-\!7) \xi^2 \right)
\right.
\hspace{10mm}
\nonumber \\
+\xi \left( (n-4) \xi+4 \right)
(n-1) {\cal{K}} {\cal{Q}} \left( p_1^2 (p_1 p_3)+p_2^2 (p_2 p_3) \right)
\hspace{27mm}
\nonumber \\
\left. \left.
-\xi \left( (n-3) \xi+2 \right)
(n-1) p_1^2 p_2^2 p_3^2 (p_1^2-p_2^2)^2
\right] \frac{}{} \right\} ,
\hspace{17mm}
\end{eqnarray}
\begin{equation}
\label{S(1,xi)}
S^{(1, \xi)}(p_1^2, p_2^2, p_3^2) = 0 ,
\hspace{80mm}
\end{equation}
\begin{eqnarray}
\label{F(1,xi)}
F^{(1, \xi)}(p_1^2, p_2^2; p_3^2)
= \frac{g^2 \; \eta}{(4\pi)^{n/2}}\; C_A \; 
\frac{1}{32 (n-1) {\cal{K}}^3 \; p_1^2 p_2^2 p_3^2}
\hspace{62mm}
\nonumber \\
\times\! \left\{
2 \left[ p_1^2 p_2^2 p_3^2
  \left[ (p_1 p_2) (p_1 p_3) (p_2 p_3) (n-1)
\right. \right. \right.
\hspace{82.9mm}
\nonumber \\
\times \left( 8 (n+1)+8 (n-3) \xi+(3 n^2-38 n+63) \xi^2
                  -(n-3) (7 n-13) \xi^3 \right)
\nonumber \\
-2 {\cal{K}} {\cal{Q}} \left( 8 (n-2)+8 (n-1) \xi+(n-1) (3 n-14) \xi^2
                                -2 (n-1) (n-3) \xi^3 \right)
\nonumber \\
\left.
+{\cal{K}} (p_1 p_2)
 \left( 8 (n+3)-(n-1) (n^2-4 n+23) \xi^2-7 (n-1) (n-3) \xi^3 \right)
\right]
\hspace{6mm}
\nonumber \\
     -\xi \left( (n-4) \xi+4 \right) (n-1)
        \left[ ((n-3) \xi+2) (n-3) (p_1 p_2)^2 (p_1 p_3)^2 (p_2 p_3)^2
\right.
\hspace{20mm}
\nonumber \\
+ \left( (n-3) \xi+n-1 \right) {\cal{K}} (p_1 p_2)^2 (p_1 p_3) (p_2 p_3)
\hspace{22mm}
\nonumber \\
\left. \left.
- \left( (n-3) \xi+3 n-7 \right)
{\cal{K}}^2 p_1^2 p_2^2+3 (n-4) {\cal{K}}^3 \right] \right]
\hspace{14mm}
\nonumber \\
\times \left[ p_3^2 (p_1 p_2) \varphi+(p_1 p_3) \kappa_1 
             +(p_2 p_3) \kappa_2+p_3^2 \kappa_3 \right]
\nonumber \\
+2 {\cal{K}} \left[ p_1^2 p_2^2 p_3^2
\right.
\hspace{128mm}
\nonumber \\
\times\! \left[ (p_1 p_3) (p_2 p_3) (n\!-\!4)
\left( 8 (n\!+\!1)+8 (n\!-\!1) \xi
      +(n\!-\!1) (3 n\!-\!23) \xi^2-5 (n\!-\!1) (n\!-\!3) \xi^3 \right)
\right.
\hspace{1mm}
\nonumber \\
\left.
 -{\cal{K}} \left( 8 (5 n\!-\!11)+4 (n\!-\!1) (5 n\!-\!11) \xi
   +(n\!-\!1) (n^2 \!+\!4 n\!-\!10) \xi^2
   +n (n\!-\!1) (n\!-\!3) \xi^3 \right) \right]
\nonumber \\
  -\xi \left( (n-4) \xi+4 \right) (n-1) (n-4)
\hspace{90mm}
\nonumber \\
\left.
 \times \left[ \left( (n-3) \xi+2 \right) (p_1 p_2) (p_1 p_3)^2 (p_2 p_3)^2
 +{\cal{K}} (p_1 p_2) (p_1 p_3) (p_2 p_3)-3 {\cal{K}}^2 (p_1 p_2) \right]
\right]
\nonumber \\
 \times \left[ (p_1 p_2) \varphi+\kappa_3 \right]
\nonumber \\
+ 2 {\cal{K}} \varphi
\left[ p_1^2 p_2^2 (p_3^2)^2
\left[ {\cal{K}} (n-1) \left( 16+16 \xi+(5 n-32) \xi^2-6 (n-3) \xi^3 \right)
\right. \right.
\hspace{30mm}
\nonumber \\
\left.
 +p_3^2 (p_1 p_2) \left(8 (n\!+\!1)\!+\!8(n\!-\!1)\xi
\!+\!(n\!-\!1)(3 n\!-\!23)\xi^2
\!-\!5(n\!-\!1)(n\!-\!3)\xi^3 \right) \right]
\nonumber \\
-\xi \left( (n-4) \xi+4 \right) (n-1) (p_1 p_3) (p_2 p_3)
\hspace{75mm}
\nonumber \\
\times\!
 \left[ \left( (n\!-\!3)\xi\!+\!2 \right) p_3^2 (p_1 p_2) (p_1 p_3) (p_2 p_3)
\left.
+\left( (n\!-\!3)\xi\!+\!3 \right) {\cal{K}} p_3^2 (p_1 p_2)
               \!+\!\xi(n-4){\cal{K}}^2 \right] \right]
\nonumber \\
     +{\cal{K}}^2 p_1^2 p_2^2 p_3^2
  \left( 8 (4 n\!-\!7)\!+\!4 (n\!-\!1) (5 n\!-\!11) \xi
         \!+\!(n\!-\!1) (13 n\!-\!30) \xi^2
         \!+\!2 (n\!-\!1) (n\!-\!2) (n\!-\!3) \xi^3 \right)
\hspace*{-4mm}
\nonumber \\
\times
\left[ \kappa_1 + \kappa_2 \right]
\nonumber \\
+{\cal{K}} \frac{\kappa_1\!-\!\kappa_2}{p_1^2\!-\!p_2^2} \!
 \left[ p_1^2 p_2^2 p_3^2
 \left[ {\cal{K}} (p_3^2\!-\!4 (p_1 p_2))
   \left( 8 (4 n\!-\!3)\!+\! 4 (n\!-\!1) (5 n\!-\!17) \xi
   \!-\!(n\!-\!1) (3 n\!-\!10) \xi^2 \right)
\right. \right.
\nonumber \\
\nonumber \\
\left.
+2 (p_1 p_2) (p_1^2-p_2^2)^2 (n+1)
   \left( 8-(n-1) \xi^2 \left( (n-3) \xi+3 \right) \right) \right]
\hspace{15mm}
\nonumber \\
-2 \xi \left( (n-4) \xi+4 \right) (n-1) {\cal{K}}
\hspace{81mm}
\nonumber \\
\left. \left.
\times \left[
{\cal{K}} \left( p_1^2 (p_1 p_3)+p_2^2 (p_2 p_3) \right)
-\left( (n-3) \xi+2 \right) (p_1 p_3) (p_2 p_3) (p_1^2-p_2^2)^2
\right] \right]
\frac{}{} \right\} ,
\hspace{4mm}
\end{eqnarray}
\begin{eqnarray}
\label{H(1,xi)}
H^{(1, \xi)}(p_1^2, p_2^2, p_3^2)
= \frac{g^2 \; \eta}{(4\pi)^{n/2}}\; C_A \; 
\frac{1}{16 {\cal{K}}^3 p_1^2 p_2^2 p_3^2}
\hspace{75mm}
\nonumber \\
\times\!\! \left\{ \left[
p_1^2 p_2^2 p_3^2 (p_1 p_2) (p_1 p_3) (p_2 p_3)
\left( 8(n\!+\!1)\!+\!14(n\!-\!1)\xi\!+\!4(n\!-\!1)(n\!-\!7)\xi^2
\!-\!5(n\!-\!1)(n\!-\!3)\xi^3 \right)
\right. \right.
\hspace*{-4mm}
\nonumber \\
-\xi \left( (n-4) \xi+4 \right)
\left[ (n-1) \left( (n-3) \xi+3 \right) (p_1 p_2)^2 (p_1 p_3)^2 (p_2 p_3)^2
                               +3 (n-4) {\cal{K}}^3 \right]
\hspace{2mm}
\nonumber \\
\left.
-(n-1) \xi (2-\xi) (p_1^2)^2 (p_2^2)^2 (p_3^2)^2 \right]
\hspace{80mm}
\nonumber \\
\times \left[ (p_1 p_2) (p_1 p_3) (p_2 p_3) \varphi
             +(p_1 p_2) (p_1 p_3) \kappa_1
             +(p_1 p_2) (p_2 p_3) \kappa_2
             +(p_1 p_3) (p_2 p_3) \kappa_3 \right]
\nonumber \\
+{\cal{K}} \left[ -p_1^2 p_2^2 p_3^2 (p_1 p_2) (p_1 p_3) (p_2 p_3)
                \left( 24+38 \xi+4 (3n-16) \xi^2-9 (n-3) \xi^3 \right)
\right.
\hspace{20mm}
\nonumber \\
\left.
 +3 \xi \left( (n\!-\!4)\xi\!+\!4 \right) \left( (n\!-\!3)\xi\!+\!3 \right)
                      (p_1 p_2)^2 (p_1 p_3)^2 (p_2 p_3)^2
 +\xi(2\!-\!\xi) (p_1^2)^2 (p_2^2)^2 (p_3^2)^2 \right]
\hspace{1mm}
\nonumber \\
\times \left[ {\cal{Q}} \varphi+\kappa_1+\kappa_2+\kappa_3 \right]
\nonumber \\
+{\cal{K}}^3 \; \varphi \; \left[ p_1^2 p_2^2 p_3^2
      \left( 16+4 (3 n-8) \xi-(n-2) \xi^2-(n-2) \xi^3 \right)
\right.
\hspace{46mm}
\nonumber \\
\left.
-\xi^2 \left( (n-4) \xi+4 \right)
  (n-4) (p_1 p_2) (p_1 p_3) (p_2 p_3) \right]  
\hspace{30mm}
\nonumber \\
+\frac{{\cal{K}}}{n-1}
\left[ 2 p_1^2 p_2^2 p_3^2 \left( 4 (n-2)+6(n-1) \xi
           +(n-1) (2 n-9) \xi^2-(n-1) (n-3) \xi^3 \right)
\right.
\hspace{5mm}
\nonumber \\
\left.
-\xi \left( (n-4) \xi+4 \right) \left( (n-3) \xi+3 \right) (n-1)
                     (p_1 p_2) (p_1 p_3) (p_2 p_3) \right]
\hspace{29mm}
\nonumber \\
\times \left[ 
p_1^2 \left( p_1^2 (p_2 p_3)+(p_1 p_2) (p_1 p_3) \right) \kappa_1
+p_2^2 \left( p_2^2 (p_1 p_3)+(p_1 p_2) (p_2 p_3) \right) \kappa_2
\right.
\hspace{8mm}
\nonumber \\
\left. \left.
+p_3^2 \left( p_3^2 (p_1 p_2)+(p_1 p_3) (p_2 p_3) \right) \kappa_3
\right] \right\} .
\hspace{5mm}
\end{eqnarray}

One of the main technical problems we met in this calculation
was how to bring the results for arbitrary $\xi$ to a reasonably
short form. Original {\sf REDUCE} output  for the numerators of
the scalar functions
(\ref{BC-ggg}) was  really huge. Then, the problem was how to organize
the result and which bases to choose. The first basis we needed was
one in the ``space'' of the scalar functions $\varphi$ and $\kappa_i$.
It was possible to get better factorization of the coefficients
by considering not these functions themselves but certain linear
combinations. Moreover, not all ``convenient'' combinations
happened to be the same for the different functions, 
see eqs.~(\ref{A(1,xi)})--(\ref{H(1,xi)}). 
Then, the coefficients multiplying these combinations of $\varphi$ and
$\kappa_i$ are polynomials in $\xi, \; n$ and the momentum invariants.
Trying to write the latter only as $p_1^2, p_2^2, p_3^2$, we were still
getting rather long expressions. The next idea was
to try to use in some cases also the scalar products $(p_1 p_2), (p_1 p_3)$
and $(p_2 p_3)$, together with the notation (\ref{qqq}) and (\ref{kkk})
for symmetric combinations. These tricks (as well as 
looking for proper combinations of $\xi$ and $n$) allowed us to
write the expressions in a much shorter form. However,
this part of the work could not be completely automatized since 
$p_i^2$ and the scalar products $(p_i p_j)$ are linearly dependent. 

There are some special values of the gauge parameter $\xi$
we would like to point out.
First of all, we see that the terms containing
$p_1^2, p_2^2$ or $p_3^2$ in the denominator disappear not only if we
put $\xi=0$ (Feynman gauge), but also in a ``singular'' (in four dimensions)
gauge\footnote{It is not clear whether the
second choice could be of use in realistic calculations, since
singularity of $\xi$ in four dimensions requires extra care in
renormalizing, etc.}, $\xi=-4/(n-4)$. 
Having no $p_i^2$ in the denominator is especially convenient 
when one considers on-shell limits, i.e., when some of the
external momenta squared vanish;
otherwise, one needs to expand the scalar integrals
in the vanishing momenta squared (see Section~4).
Secondly, many terms vanish for $\xi=-2/(n-3)$, which could be
considered an $n$-dimensional generalization of the Fried--Yennie
gauge \cite{FriedYennie} (see also ref.~\cite{AbrSol} and Appendix D).

\subsection{Contributions of the quark loops}

Let us consider the quark loop contributions to the functions
(\ref{BC-ggg}). We assume that there are $N_f$ quarks which are all 
massless, and we define
\begin{equation}
\label{T_R}
T_R = {\textstyle{1\over8}} \; \mbox{Tr}(I) = {\textstyle{1\over2}}
\hspace{10mm} (\mbox{if} \;\;\;\; \mbox{Tr}(I)=4 ) ,
\end{equation}  
where $I$ is the ``unity'' in the space of Dirac matrices.

The quark loop contributions do not depend on $\xi$. 
The results of the calculation are:
\begin{equation}
\label{A(1,q)}
A^{(1,q)}(p_1^2, p_2^2; p_3^2)
= \frac{g^2 \; \eta}{(4\pi)^{n/2}} \; 
N_f T_R  \; \frac{n-2}{n-1} \; \left[ \kappa_1 + \kappa_2 \right] ,
\end{equation}
\begin{equation}
B^{(1,q)}(p_1^2, p_2^2; p_3^2)
=   \frac{g^2 \; \eta}{(4\pi)^{n/2}} \;
N_f T_R \; \frac{n-2}{n-1} \; \left[ \kappa_1 - \kappa_2 \right] ,
\end{equation}
\begin{equation}
C^{(1,q)}(p_1^2, p_2^2; p_3^2)
= \frac{g^2 \; \eta}{(4\pi)^{n/2}} \; 
2 N_f T_R \; \frac{n-2}{n-1} \; \frac{\kappa_1-\kappa_2}{p_1^2 - p_2^2} ,
\end{equation}
\begin{equation}
\label{S(1,q)}
S^{(1,q)}(p_1^2, p_2^2, p_3^2) = 0 ,
\end{equation}
\begin{eqnarray}
F^{(1,q)}(p_1^2, p_2^2; p_3^2) = - \frac{g^2 \; \eta}{(4\pi)^{n/2}} \; 
N_f T_R \; 
\frac{1}{(n-1)(n-2) {\cal{K}}^3}
\hspace{58mm}
\nonumber \\
\times \! \left\{ 2  \left[(n^2-1) (p_1 p_2) (p_1 p_3) (p_2 p_3)
              +2 (n-2) p_3^2 {\cal{K}}-(n-7) (p_1 p_2) {\cal{K}}\right]
\right.
\hspace{34mm}
\nonumber \\
\times \left[ p_3^2 (p_1 p_2) \varphi + (p_1 p_3) \kappa_1
              +(p_2 p_3) \kappa_2 + p_3^2 \kappa_3 \right]
\hspace{5mm}
\nonumber \\
       +2 (n+1)(n-4) {\cal{K}} \; p_3^2 (p_1 p_2) \;
\left[(p_1 p_2) \varphi+\kappa_3\right]
\hspace{68mm}
\nonumber \\
       +2 p_3^2 {\cal{K}} \; \varphi \; 
\left[(n+1) (p_1 p_3) (p_2 p_3) +(n-3) {\cal{K}}\right] 
+ n (n-4)
{\cal{K}}^2  \left[ \kappa_1+\kappa_2 \right]
\hspace{25mm}
\nonumber \\
\left.
+  {\cal{K}} \; \frac{\kappa_1-\kappa_2}{p_1^2 - p_2^2} \;
\left[ 2 (n+1) (p_1 p_2) (p_1^2 - p_2^2)^2
       + (n-2)^2 {\cal{K}}
         \left( p_1^2 + p_2^2 - 2 (p_1 p_2) \right)
\right]
\right\} ,
\hspace{10mm}
\end{eqnarray}
\begin{eqnarray}
\label{H(1,q)}
H^{(1,q)}(p_1^2, p_2^2, p_3^2)
= - \frac{g^2 \; \eta}{(4\pi)^{n/2}} \; 
2 N_f T_R \; \frac{1}{(n-1)(n-2){\cal{K}}^3} \;
\hspace{50mm}
\nonumber \\
\times
\left\{ (n^2 -1) (p_1 p_2)(p_1 p_3)(p_2 p_3)
\right. 
\hspace{93mm}
\nonumber \\
\times
\left[ (p_1 p_2)(p_1 p_3)(p_2 p_3) \varphi + (p_1 p_2)(p_1 p_3) \kappa_1 
+ (p_1 p_2)(p_2 p_3) \kappa_2 + (p_1 p_3)(p_2 p_3) \kappa_3 \right]
\nonumber \\
- 3 (n-1) (p_1 p_2)(p_1 p_3)(p_2 p_3) {\cal{K}}
\left[ {\cal{Q}} \varphi +  \kappa_1 +  \kappa_2 + \kappa_3 \right]
+ (n-1)(n-2) \; {\cal{K}}^3 \; \varphi
\hspace{10mm}
\nonumber \\
+ (n-2) {\cal{K}}
\left[
p_1^2 \left( p_1^2 (p_2 p_3) + (p_1 p_2)(p_1 p_3) \right) \; \kappa_1
+ p_2^2 \left( p_2^2 (p_1 p_3) + (p_1 p_2)(p_2 p_3) \right) \; \kappa_2
\right.
\nonumber \\
\left.  \left.
+ p_3^2 \left( p_3^2 (p_1 p_2) + (p_1 p_3)(p_2 p_3) \right) \; \kappa_3
\right] \right\} .
\hspace{10mm}
\end{eqnarray}

\subsection{Symmetric limit and comparison}

Now, we would like compare our results with those
by Celmaster and Gonsalves \cite{CG}.
For their study of renormalization-prescription dependence of
Green's functions, they evaluated the three-gluon
vertex function to one loop at the symmetric point,
\begin{equation}
p_1^2=p_2^2=p_3^2 \equiv p^2 = -M^2 .
\end{equation}
In this case, we get 
$(p_1 p_2)=(p_1 p_3)=(p_2 p_3)=-{\textstyle{1\over2}} p^2 
= {\textstyle{1\over2}}M^2$.
At the symmetric point, the vertex function simplifies
considerably.
First of all, because of their antisymmetry, the $B$ and $S$
functions (\ref{BC-ggg}) must be zero,
\begin{equation}
B(p^2, p^2; p^2) = S(p^2, p^2, p^2) \equiv 0 .
\end{equation}
Furthermore, in this limit, the number of independent tensor
combinations in the three-gluon vertex reduces to three,
and the vertex function can be written, in the notation used in \cite{CG},
as\footnote{The counterterm contribution is omitted.}
\begin{eqnarray}
\Gamma_{\mu_1\mu_2\mu_3}(p_1,p_2,p_3)
= G_0(p^2) \left[ g_{\mu_1\mu_2}(p_1-p_2)_{\mu_3}
                  + g_{\mu_2\mu_3}(p_2-p_3)_{\mu_1}
                  + g_{\mu_3\mu_1}(p_3-p_1)_{\mu_2} \right]
\nonumber \\
- G_1(p^2) (p_2-p_3)_{\mu_1} (p_3-p_1)_{\mu_2} (p_1-p_2)_{\mu_3}
\hspace{27mm}
\nonumber \\
+ G_2(p^2)
\left( {p_1}_{\mu_3} {p_2}_{\mu_1} {p_3}_{\mu_2}
-  {p_1}_{\mu_2} {p_2}_{\mu_3} {p_3}_{\mu_1} \right) ,
\hspace{30mm}
\end{eqnarray}
with the three $G_i$ functions related to
the scalar functions in (\ref{BC-ggg}) through
\begin{eqnarray}
G_0(p^2)
&=& A(p^2,p^2;p^2) + \textstyle{1\over2} p^2 C(p^2,p^2;p^2)
+ \textstyle{1\over4} (p^2)^2 F(p^2,p^2;p^2)
+ \textstyle{1\over2}  p^2 H(p^2,p^2,p^2) , 
\hspace*{8mm}
 \\
G_1(p^2)
&=&C(p^2,p^2;p^2) + \textstyle{1\over2}  p^2 F(p^2,p^2;p^2) ,
\\
G_2(p^2)
&=&C(p^2,p^2;p^2) + \textstyle{1\over2}  p^2 F(p^2,p^2;p^2)
+H(p^2,p^2,p^2) .
\end{eqnarray}
We note that two of these relations may be expressed
more compactly as
\begin{equation}
G_2(p^2)
=G_1(p^2)+H(p^2,p^2,p^2) , 
\hspace{5mm}
G_0(p^2)
=A(p^2,p^2;p^2)+\textstyle{1\over2}  p^2 G_2(p^2) .
\end{equation}

From our results, we obtain the following
expressions for the $G_i$ functions (in arbitrary gauge and dimension):
\begin{eqnarray}
G_0^{(1,\xi)}(p^2) = \frac{g^2 \; \eta}{(4\pi)^{n/2}} \;
C_A \; \frac{1}{288}
\hspace{93mm}
\nonumber \\
\times
\left\{ p^2 \varphi
\left[ 8 + 12 \xi (14 n-51) + 6\xi^2 (n^2-18n+60) - \xi^3 (n-4)(n-12)
\right]
\right.
\hspace{10mm}
\nonumber \\
\left.
- 6 \kappa
\left[ 32 + 36\xi (2n-7) + 6\xi^2 (n-4)(n-6) - \xi^3 (n-4)(n-3)
\right]
\right\} ,
\hspace{5mm}
\end{eqnarray}
\begin{eqnarray}
G_1^{(1,\xi)}(p^2) = -  \frac{g^2 \eta}{(4\pi)^{n/2}} \;
C_A \; \frac{1}{3456 \; p^2}
\hspace{89mm}
\nonumber \\
\times \!\!
\left\{ p^2 \varphi
\left[ 64(n\!-\!20) \!-\! 144 \xi (7n\!-\!26)
\!-\! 24\xi^2 (n^2\!-\!18n\!+\!50)
\!+\! \xi^3 (n^3 \!-\! 24 n^2 \!+\! 200n \!-\! 384)
\right] 
\right.
\hspace{2mm}
\nonumber \\
\left.
- 6 \kappa \frac{n-4}{n-1} 
\left[ 64(n\!-\!2) \!-\! 144\xi (n\!-\!1) \!-\! 12\xi^2 (n\!-\!1)(2n\!-\!7)
\!+\! \xi^3 (n\!-\!1)(n\!-\!3)(n\!-\!20)
\right]
\right\} ,
\hspace{3mm}
\end{eqnarray}
\begin{eqnarray}
G_2^{(1, \xi)}(p^2) = \frac{g^2 \eta}{(4\pi)^{n/2}} \;
C_A \; \frac{1}{144 \; p^2}
\hspace{86mm}
\nonumber \\
\times
\left\{ p^2 \varphi
\left[ 128 + 6 \xi (29n\!-\!114) + 6\xi^2 (n^2\!-\!18n\!+\!60)
- \xi^3 (n \!-\! 4)(n \!-\! 12)
\right]
\right.
\hspace{10mm}
\nonumber \\
\left.
- 6 \kappa \frac{n-4}{n-1} \;
\left[ 8 + 30\xi (n\!-\!1) + 6\xi^2 (n\!-\!1)(n\!-\!5)
- \xi^3 (n\!-\!1)(n\!-\!3)
\right]
\right\} ,
\hspace{12mm}
\end{eqnarray}
\begin{equation}
G_0^{(1,q)}(p^2) = -\frac{g^2 \; \eta}{(4\pi)^{n/2}} \; N_f T_R \; 
\frac{2(3n-8)}{9(n-2)}
\; \left\{ 2 p^2 \; \varphi - 3 \; \kappa \right\} ,
\end{equation}
\begin{equation}
G_1^{(1,q)}(p^2) =  - \frac{g^2 \; \eta}{(4\pi)^{n/2}} \;
N_f T_R \; \frac{4}{27\; p^2}
\; \left\{ 4 p^2 \; \varphi
+ 3 \kappa \; \frac{n-4}{n-1} \right\} ,
\end{equation}
\begin{equation}
G_2^{(1,q)}(p^2) = - \frac{g^2 \; \eta}{(4\pi)^{n/2}} \;
N_f T_R \; \frac{4}{9 (n-2)\; p^2}
\; \left\{ 2 p^2 \; \varphi\; (3n-8)
- 3 \kappa \; \frac{n-4}{n-1}  \right\} ,
\end{equation}
where $\varphi = \varphi(p^2, p^2, p^2)$
and $\kappa = \kappa(p^2)$.
Expanding these results around $n=4$ and keeping the divergent
and finite (in $\varepsilon=(4-n)/2$) terms only, we arrive at {\em exactly}
the same results as Celmaster and Gonsalves, see eqs.~(14a), (14b)
and (14c) of \cite{CG}\footnote{Their transcendental constant $I$ 
is nothing but our $\left. p^2 \varphi(p^2,p^2,p^2)\right|_{n=4}$ 
(see eq.~(\ref{Cl2(pi/3)}) )
which can be expressed in terms of Clausen's function as
$\textstyle{4\over{\sqrt{3}}} 
\mbox{Cl}_{2}\left(\textstyle{\pi\over3}\right)$.}.

In ref.~\cite{Vendr}
the QCD renormalization has been considered at an asymmetric point,
\begin{equation}
\label{asymmetric}
p_1^2=p_2^2 \equiv p^2 = -M^2, 
\hspace{10mm} 
p_3^2 = 4 z p^2 = -4z M^2 .
\end{equation}
In particular, the three-gluon vertex was studied at this point,
including quark loop contributions (with massive quarks).
The tensor structures used to decompose the three-gluon vertex
are presented in eqs.~(2) and (4) of \cite{Vendr}.
Of the scalar functions multiplying the seven tensor structures
defined by eq.~(5) of \cite{Vendr}, an explicit result is presented
for the function $F_0$ only, see eq.~(6) of \cite{Vendr}\footnote{
We note some misprints in \cite{Vendr}:
(i) in the definition of the $\varphi$ function (eq.~(8)),
$x$ in the denominator of the expression under the square root
should read $z$;
(ii) before eq.~(7), the definition of $1/\hat{\varepsilon}$ should read
$1/\hat{\varepsilon} = 2/(D-4)+\gamma+\ln(Q^2/(4\pi\nu^2))$
(the sign before the logarithm should be changed).}.
In terms of the scalar functions (\ref{BC-ggg}), one finds 
the correspondence
$F_0 \leftrightarrow - A(p^2, 4zp^2; p^2) - B(p^2, 4zp^2; p^2)$.

Calculating this combination of the $A$ and $B$ functions for the
case (\ref{asymmetric}), we find coincidence of the $\xi$ and 
$\xi^2$ contributions\footnote{In \cite{Vendr}, 
$b$ is the same as our $\xi$. 
The integral $I(0,z)$ from \cite{Vendr} is
related, in the limit (\ref{asymmetric}), to our $J(1,1,1)$ as
$I(0,z) = - (i\pi^2)^{-1} M^2 J(1,1,1)$.}, as well as the quark loop
contibutions in the massless limit. However, our result for
the Feynman-gauge part is different. So, we do {\em not} confirm 
eq.~(6) of \cite{Vendr}. 

\subsection {Renormalization}

In the limit $n\to 4$ ($\varepsilon\to 0$), the only 
function which may have an ultraviolet singularity is the $A$ function, 
since this is the only function which does not vanish at the 
``zero-loop'' level, see eq.~(\ref{A0=1}). In arbitrary gauge, the 
ultraviolet-singular 
part of the $A$ function follows from eq.~(\ref{A(1,xi)}),
\begin{equation}
\label{A_UV}
A^{(1, UV)} = \frac{g^2 \; \eta}{(4\pi)^{2-\varepsilon}} \;
\left[ - C_A 
\left({\textstyle{2\over3}}+ {\textstyle{3\over4}} \xi \right)
+ {\textstyle{4\over3}} N_f T_R \right] \; \kappa^{(UV)} ,
\end{equation}
where $\kappa^{(UV)}=1/\varepsilon+\ldots$ is the ultraviolet-singular
part of the $\kappa$ function (\ref{kappa}).
Therefore, the divergent part of the counterterm contribution 
should be equal to minus the r.h.s.\ of eq.~(\ref{A_UV}).
This counterterm contribution can be written
as\footnote{In commonly-used notation, the expression (\ref{A_CT}) 
corresponds to $(Z_1-1)$ at the one-loop order.}
\begin{eqnarray}
\label{A_CT}
A^{(1,CT)} &=& \frac{{\overline{g}}^2}{(4\pi)^2} \;
\left[
C_A \left( {\textstyle{2\over3}} + {\textstyle{3\over4}} \xi \right)
- {\textstyle{3\over4}} N_f T_R \right]
\left( \frac{1}{\varepsilon} + R \right)
\nonumber \\
&=& \frac{g^2 \; \eta}{(4\pi)^{2-\varepsilon}} \;
\left[
C_A \left( {\textstyle{2\over3}} + {\textstyle{3\over4}} \xi \right)
- {\textstyle{3\over4}} N_f T_R \right]
\left( \frac{1}{\varepsilon} + R \right) + {\cal{O}}(\varepsilon) ,
\end{eqnarray}
where $R$ is a constant corresponding to the choice of renormalization
scheme, whereas
${\overline{g}}^2 \equiv g^2 e^{-\gamma \varepsilon} (4\pi)^{\varepsilon}
= g^2 \exp\{ \varepsilon (-\gamma+\ln(4\pi) ) \} $ is the ``rescaled''
coupling constant. Such a re-definition of $g^2$ is usually
performed in the context of the ${\overline{\mbox{MS}}}$
renormalization scheme \cite{MSbar} which corresponds to the choice 
$R=0$ (for $R=0$, eq.~(\ref{A_CT}) corresponds to 
eq.~(15) of ref.~\cite{CG}).
The second line of eq.~(\ref{A_CT}) is more convenient for dealing with
the expressions obtained in the present paper, since one can
keep $g^2 \eta/(4\pi)^{n/2}$ as an overall factor. Here, we have
used the fact that
\begin{equation}
\label{eta-exp}
\eta = e^{-\gamma \varepsilon} \; \left( 1 - {\textstyle{1\over{12}}}
\pi^2 \varepsilon^2 + {\cal{O}}(\varepsilon^3) \right) .
\end{equation}
The $\varepsilon^2$ term in (\ref{eta-exp}) is not relevant for
the ultraviolet renormalization at the {\em one}-loop level.
However, it yields finite contributions when one has
$1/\varepsilon^2$ infrared (on-shell) singularities (see section~4.3).
              
If we now recall the existence of the dimensional-regularization 
scale parameter $\mu_{DR}$ (which we usually put equal to one, see 
footnote at the end of section~2),
we shall see that, as opposed to other one-loop-order contributions,  
the counterterm (\ref{A_CT}) should {\em not} be
multiplied by the factor $(\mu_{DR})^{2\varepsilon}$, 
see also in ref.~\cite{CG}.

Examining eq.~(\ref{A_UV}), it is interesting to note that for
\begin{equation}
\left. \frac{}{} \xi \right|_{n=4} \equiv \xi_0 
= {\textstyle{8\over9}} \left( \frac{2N_f T_R}{C_A} -1 \right)
\end{equation}
we do not have any ultraviolet singularity in the one-loop contribution
to the three-gluon vertex. If we substitute $N_f=6, \; C_A=3$ and
$T_R={\textstyle{1\over2}}$, we get $\xi_0={\textstyle{8\over9}}$.

\section{On-shell limits}
\setcounter{equation}{0}

There are two main on-shell cases of interest: when one or
two of the external momenta squared are zero.
However, it is also instructive to consider, as a separate case, 
the limit when one external momentum (not only its square) 
vanishes.

\subsection{One external momentum squared is zero, $p_3^2=0$}

In this case, we should substitute in the expressions for all
scalar functions 
\begin{equation}
\label{p_3^2=0}
p_3^2 = 0 , \;\; (p_1 p_2) = -\textstyle{1\over2}(p_1^2 + p_2^2) , \;\;
(p_1 p_3) = -\textstyle{1\over2}(p_1^2-p_2^2) , \;\;
(p_2 p_3) = \textstyle{1\over2}(p_1^2-p_2^2) .
\end{equation}
Note that now we should consider the scalar functions $A, B, C$ 
and $F$  from (\ref{BC-ggg}) with permuted arguments as well. 

The result for the triangle integral (\ref{J(1,1,1)}) 
simplifies in this limit, 
\begin{equation}
\label{J111_p3^2=0}
\left. \frac{}{} J(1,1,1) \right|_{p_3^2=0} 
= \mbox{i} \pi^{2-\varepsilon} \; \eta \; \varphi(p_1^2,p_2^2,0)  
= \mbox{i} \pi^{2-\varepsilon}
\; \eta \; 
\; \frac{1}{\varepsilon^2} \;
\frac{(-p_1^2)^{-\varepsilon} - (-p_2^2)^{-\varepsilon}}{p_1^2 - p_2^2} ,
\end{equation}
where $\eta$ is defined by eq.~(\ref{eta}). Moreover,
in the framework of dimensional regularization, \cite{dimreg},
\begin{equation}
\label{J(1,1,0)=0}
\left. \frac{}{} J(1,1,0) \right|_{p_3^2=0} = 0 ,
\end{equation}
while the results for $J(1,0,1)$ and $J(0,1,1)$ remain unchanged.
As to the factor $1/\varepsilon^2$ in (\ref{J111_p3^2=0}), one power of
$\varepsilon$ is cancelled by the expansion of the numerator, while
another power of $\varepsilon$ survives and corresponds to the infrared
(on-shell) singularity which arises in the scalar integral in the 
limit $p_3^2=0$.

For the Feynman gauge, $\xi=0$ (and also for the singular gauge,
$\xi=-4/(n-4)$), it is enough to perform the above substitutions
to get the answer. In the case of arbitrary $\xi$, however, the 
situation is more tricky, due to the presence of $p_3^2$ in the
denominators of the scalar functions. 
Here, in order to get a correct answer, one needs to expand
the integral $J(1,1,1)$ in $p_3^2$ and keep the term of 
order $p_3^2$,
\begin{eqnarray}
\label{J111_p3^2_to_0}
\left. \frac{}{} J(1,1,1) \right|_{p_3^2 \to 0} =
\mbox{i} \; \pi^{2-\varepsilon}
\; \eta
\; \frac{1}{\varepsilon^2} \;
\left\{
\frac{(-p_1^2)^{-\varepsilon} - (-p_2^2)^{-\varepsilon}}{p_1^2 - p_2^2}
\right.
\hspace{61.5mm}
\nonumber \\
\left.
-\frac{p_3^2}{(1\!+\!\varepsilon)(p_1^2 \!-\! p_2^2)^2}
\left[ (1\!-\!\varepsilon) \left( (-p_1^2)^{-\varepsilon} 
\!+\! (-p_2^2)^{-\varepsilon} \right)
+\!2 \frac{(-p_1^2)^{1-\varepsilon}
\!-\!(-p_2^2)^{1-\varepsilon}}{p_1^2 - p_2^2} \right]
\right\}
+ \!{\cal{O}}\!\left((p_3^2)^2\right)\!.
\hspace{1.9mm}
\end{eqnarray}

To present the results obtained for the scalar functions 
(\ref{BC-ggg}) in this limit, it is convenient to introduce
\begin{equation}
\label{delta12}
\delta_{12} \equiv \frac{p_1^2-p_2^2}{p_1^2+p_2^2} \; .
\end{equation}
In this section, we present such results for the three-gluon
scalar functions in the Feynman gauge, and also for the quark loop
contributions. The expressions for arbitrary covariant gauge
are listed in Appendix~E.
We present the results for the $A, B, C$ and $F$ functions 
of the arguments $(p_1^2,p_2^2;0)$ and $(0,p_1^2;p_2^2)$. 
The results for the third set of arguments corresponding to the 
cyclic permutations in eq.~(\ref{BC-ggg}), $(p_2^2,0;p_1^2)$,
can be obtained from the functions of $(0,p_1^2;p_2^2)$ by
using the symmetry (for the $A, C$ and $F$ functions) or
antisymmetry (for the $B$ function) with respect to the first
two arguments, and interchanging $p_1^2 \leftrightarrow p_2^2$.
The $H$ function with permuted arguments does not change, while the
$S$ function is zero (at one loop).

The results  for the gluon and ghost contributions to the three-gluon
scalar functions in the Feynman gauge are:
\begin{eqnarray}
\label{A10(p1,p2;0)}
A^{(1, 0)}(p_1^2, p_2^2; 0)
= - \frac{g^2 \; \eta}{(4\pi)^{n/2}} \; C_A \; \frac{1}{4 (n-1) (n-4)}
\hspace{67mm}
\nonumber \\
\times
\left\{ (n-4) (3n-2) \left[ \kappa_1 + \kappa_2 \right]
- n(n-1) (\delta_{12})^{-1} \left[ \kappa_1 - \kappa_2 \right]
\right\} ,
\hspace{4mm}
\end{eqnarray}
\begin{eqnarray}
A^{(1, 0)}(0, p_1^2; p_2^2)
= - \frac{g^2 \; \eta}{(4\pi)^{n/2}} \; C_A \;
\frac{1}{4 (n-1) (n-4) \; (p_1^2-p_2^2)}
\hspace{49mm}
\nonumber \\
\times
\left\{ (n-4) \kappa_1 \left( (2n-1) p_1^2 - (4n-3) p_2^2 \right)
- 2 (n-1)  (\delta_{12})^{-1} \left[ \kappa_1 - \kappa_2 \right]
\left( 2 p_1^2 - 3 p_2^2 \right)
\right\} ,
\hspace{3mm}
\end{eqnarray}
\begin{equation}
B^{(1, 0)}(p_1^2, p_2^2; 0)
= -\frac{g^2 \; \eta}{(4\pi)^{n/2}} \; C_A \;
\frac{1}{4 (n-1) (n-4)} \; (4n^2-21n+14)
\; \left[ \kappa_1 - \kappa_2 \right] ,
\end{equation}
\begin{eqnarray}
B^{(1, 0)}(0, p_1^2; p_2^2)
= -\frac{g^2 \; \eta}{(4\pi)^{n/2}} \; C_A \;
\frac{1}{4 (n-1) (n-4) \; (p_1^2-p_2^2)^2}
\hspace{47mm}
\nonumber \\
\times
\left\{ -(n-4) (p_1^2-p_2^2) \kappa_1 
\left( (6n-5) p_1^2 - (4n-3) p_2^2 \right)
\right.
\hspace{30mm}
\nonumber \\
\left.
+ 2 (n-1) p_1^2 \; \left[ \kappa_1 - \kappa_2 \right]
\left( (n-3) p_1^2 - (n-5) p_2^2 \right)
\right\} ,
\hspace{15mm}
\end{eqnarray} 
\begin{equation}
C^{(1, 0)}(p_1^2, p_2^2; 0)
= - \frac{g^2 \; \eta}{(4\pi)^{n/2}} \; C_A \;
\frac{1}{2 (n-1) \; (p_1^2-p_2^2)} \; n 
\; \left[ \kappa_1 - \kappa_2 \right] ,
\end{equation}
\begin{eqnarray}
C^{(1, 0)}(0, p_1^2; p_2^2)
= - \frac{g^2 \; \eta}{(4\pi)^{n/2}} \; C_A \;
\frac{1}{2 (n-1) (n-4) \; p_1^2 \; (p_1^2-p_2^2)^2}
\hspace{42mm}
\nonumber \\
\times \!
\left\{  (n\!-\!4)(p_1^2-p_2^2) \kappa_1 
\left( n p_1^2 - (4n\!-\!3) p_2^2 \right)
+ 6 (n\!-\!1) p_1^2 p_2^2 \left[ \kappa_1 - \kappa_2 \right] 
\right\} ,
\hspace{4mm}
\end{eqnarray}
\begin{eqnarray}
F^{(1, 0)}(p_1^2, p_2^2; 0)
= -\frac{g^2 \; \eta}{(4\pi)^{n/2}} \; C_A \;
\frac{1}{(n-1)(n-4) \; (p_1^2-p_2^2)^2}
\hspace{50mm}
\nonumber \\
\times (4n-7) \left\{ (n-4) \left[ \kappa_1 + \kappa_2 \right]
- 2 (\delta_{12})^{-1} \left[ \kappa_1 - \kappa_2 \right]
\right\} ,
\hspace{4mm}
\end{eqnarray}
\begin{eqnarray}
F^{(1, 0)}(0, p_1^2; p_2^2)
=-\frac{g^2 \; \eta}{(4\pi)^{n/2}} \; C_A \; 
\frac{1}{(n-1) (n-4) \; p_1^2 \; (p_1^2-p_2^2)^2}
\hspace{45mm}
\nonumber \\
\times \!
\left\{ (n-4)(p_1^2-p_2^2) \kappa_1
\left[ (3n+1) - (n+2) (\delta_{12})^{-1} - 6 (\delta_{12})^{-2} \right]
\right.
\hspace{24mm}
\nonumber \\
\left.
+ 2 p_1^2 \left[ \kappa_1 - \kappa_2 \right]
\left[ (n^2-12n+17) + 4(n-4) (\delta_{12})^{-1} 
+6 (\delta_{12})^{-2} \right] 
\right\} ,
\hspace{4mm}
\end{eqnarray}
\begin{eqnarray}
H^{(1, 0)}(p_1^2, p_2^2; 0)
= - \frac{g^2 \; \eta}{(4\pi)^{n/2}} \; C_A \;
\frac{1}{(n-1)(n-4) \; (p_1^2-p_2^2)}
\hspace{47mm}
\nonumber \\
\times \left\{
3 (n-4) (\delta_{12})^{-1} \left[ \kappa_1 + \kappa_2 \right] 
+ \left[ \kappa_1 - \kappa_2 \right]
\left( (n^2-2n-2) - 6 (\delta_{12})^{-2} \right)
\right\} .
\hspace{3mm}
\end{eqnarray}

The quark loop contributions in the limit $p_3^2=0$ are:
\begin{equation}
A^{(1,q)}(p_1^2, p_2^2; 0)
= \frac{g^2 \; \eta}{(4\pi)^{n/2}} \; N_f T_R \;
\frac{n-2}{n-1} \; 
\left[ \kappa_1 + \kappa_2 \right] ,
\end{equation}
\begin{equation}
A^{(1,q)}(0, p_1^2; p_2^2)
= \frac{g^2 \; \eta}{(4\pi)^{n/2}} \; N_f T_R \;
\frac{n-2}{n-1} \; \kappa_1  ,
\end{equation}
\begin{equation}
B^{(1,q)}(p_1^2, p_2^2; 0)
= \frac{g^2 \; \eta}{(4\pi)^{n/2}} \; N_f T_R \; 
\frac{n-2}{n-1} \; 
\left[ \kappa_1 - \kappa_2 \right] ,
\end{equation}
\begin{equation}
B^{(1,q)}(0, p_1^2; p_2^2)
= - \frac{g^2 \; \eta}{(4\pi)^{n/2}} \; N_f T_R \; 
\frac{n-2}{n-1} \; \kappa_1  ,
\end{equation}
\begin{equation}
C^{(1,q)}(p_1^2, p_2^2; 0)
= \frac{g^2 \; \eta}{(4\pi)^{n/2}} \; 2 N_f T_R \; 
\frac{n-2}{n-1} \; \frac{\kappa_1 - \kappa_2}{p_1^2-p_2^2} ,
\end{equation}
\begin{equation}
C^{(1,q)}(0, p_1^2; p_2^2)
= \frac{g^2 \; \eta}{(4\pi)^{n/2}} \; 2 N_f T_R \;
\frac{n-2}{n-1} \; \frac{\kappa_1}{p_1^2}  ,
\end{equation}
\begin{eqnarray}
F^{(1,q)}(p_1^2, p_2^2; 0)
= \frac{g^2 \; \eta}{(4\pi)^{n/2}} \; 4 N_f T_R \;
\frac{n}{(n-1) (n-2) \; (p_1^2-p_2^2)^2}
\hspace{28mm}
\nonumber \\
\times \left\{ (n-4) \left[ \kappa_1 + \kappa_2 \right]
- 2 (\delta_{12})^{-1} \; \left[ \kappa_1 - \kappa_2 \right] 
\right\} ,
\end{eqnarray}
\begin{eqnarray}
F^{(1,q)}(0, p_1^2; p_2^2)
= \frac{g^2 \; \eta}{(4\pi)^{n/2}} \; 4 N_f T_R \;
\frac{1}{(n-1) (n-2) (n-4) \; p_1^2 \; (p_1^2-p_2^2)}
\hspace{17mm}
\nonumber \\
\times \left\{
(n-4) \; \kappa_1 
\left[ (n^2-5n+8) - (\delta_{12})^{-1} \; (n+2) - 6 (\delta_{12})^{-2}
\right] \right.
\nonumber \\
\left.
+ 16 \frac{p_1^2 \; p_2^2}{(p_1^2-p_2^2)^3}
\left[ \kappa_1 - \kappa_2 \right] \; 
\left( (n-1) p_1^2 - (n-4) p_2^2 \right)
\right\} ,
\end{eqnarray}
\begin{eqnarray}
\label{H1q(p1,p2,0)}
H^{(1,q)}(p_1^2, p_2^2; 0)
= \frac{g^2 \; \eta}{(4\pi)^{n/2}} \; 4 N_f T_R \; 
\frac{1}{(n-1) (n-2) (n-4) \; (p_1^2-p_2^2)}
\hspace{30mm}
\nonumber \\
\times \left\{ 3(n-4) (\delta_{12})^{-1}
\left[ \kappa_1 + \kappa_2 \right]
+ \left[ (n-2)(n^2-5n+7) - 6 (\delta_{12})^{-2} \right]
 \left[ \kappa_1 - \kappa_2 \right]
\right\} .
\hspace{3mm}
\end{eqnarray}

We note that there is an interesting relation between the one-loop
contributions to the $A$ and $B$ functions of permuted 
arguments\footnote{This is also valid
for the zero-loop functions, since $A^{(0)}=1$ and $B^{(0)}=0$.},
\begin{eqnarray}
\label{A-B-relation}
(p_1^2+p_2^2) \; A^{(1)}(p_1^2,p_2^2;0)
\! &-& \! p_1^2 \; A^{(1)}(0,p_1^2;p_2^2)
         -p_2^2 \; A^{(1)}(0,p_2^2;p_1^2)
\nonumber \\
\! &+& \! p_1^2 \; B^{(1)}(0,p_2^2;p_1^2)
         +p_2^2 \; B^{(1)}(0,p_1^2;p_2^2) = 0 .
\end{eqnarray}
This relation is also satisfied by the expressions for
arbitrary $\xi$ given in Appendix~E.

The infrared $1/\varepsilon$ singularities of the results for gluon and ghost 
contributions (also in arbitrary gauge, see Appendix~E) have been compared 
with the results given in \cite{BF}, 
eqs.~(24)--(25).
The functions $G^j$ defined in \cite{BF} are proportional to our functions
$Z_{ijk}$ (see Appendix~A)
which can be represented as linear combinations of the scalar functions
(\ref{BC-ggg}), including those with permuted arguments. 
To get renormalized results, the counterterm (\ref{A_CT})
was added to all $A$ functions.  
In the ${\overline{\mbox{MS}}}$ scheme, the obtained results 
coincide\footnote{
Up to a misprint in eq.~(25g) of \cite{BF} where, in the term
proportional to $\lambda$ (their $\lambda$ is nothing but our $\xi$), 
the contribution $l(1+\textstyle{11\over4} + 2 f^2)$
should read $l(1+\textstyle{11\over4} f + 2 f^2)$.}
with those presented in ref.~\cite{BF}, eq.~(25).

\subsection{One external momentum is zero, $p_3=0$}

In this case, $p_1=-p_2\equiv p \;$ ($p_1^2=p_2^2=p^2$), 
and the proper limit of eq.~(\ref{J111_p3^2=0}) yields:
\begin{equation}
\left. \frac{}{} J(1,1,1) \right|_{p_3=0}
=  \mbox{i} \; \pi^{2-\varepsilon}\;
\; \eta 
\; \frac{1}{\varepsilon} \; (-p^2)^{-1-\varepsilon} .
\end{equation}
Actually, we get some powers of $(p_1^2-p_2^2)$ in the 
denominator from the ${\cal{K}}$'s, eq.~(\ref{kkk}), since 
${\cal{K}}=-\frac{1}{4} (p_1^2-p_2^2)^2$ in this limit. Therefore, we
should be careful taking the limit $p_2^2 \to p_1^2$
and expand the numerator up to higher powers of
\begin{equation}
\delta'_{12} \equiv \frac{p_1^2-p_2^2}{p_1^2} = 
\frac{2\delta_{12}}{1+\delta_{12}} .
\end{equation} 
Starting from the result for the case $p_3^2=0$,
we need to express $\kappa_2$ as $\kappa_1$ times an 
expansion in $\delta'_{12}$. This can easily be done using
\begin{equation}
J(1,0,1)=J(0,1,1) \left( \frac{p_2^2}{p_1^2} \right)^{-\varepsilon}
= J(0,1,1) \sum_{j=0}^{\infty} \frac{(\varepsilon)_j}{j !} (\delta'_{12})^j .
\end{equation}
In practice, we need the terms up to $(\delta'_{12})^3$ only.
It is interesting that in this special case we do not need
the $p_3^2$ term of the expansion of $J(1,1,1)$  (like in
eq.~(\ref{J111_p3^2_to_0}) )
since it cancels in all contributions.

In this limit, there are only three independent tensor structures left,
and the coefficients multiplying these structures can be expressed in terms
of the ``surviving'' scalar functions as\footnote{This corresponds 
to the decomposition used in ref.~\cite{BF}, eq.~(20).}
\begin{eqnarray}
\label{Gamma_p3=0}
\Gamma_{\mu_1 \mu_2 \mu_3}(p,-p,0)
= 2 \; g_{\mu_1 \mu_2} p_{\mu_3} \; 
\left[ A(p^2, p^2; 0) + p^2 \; C(p^2, p^2; 0) \right]
\hspace{40mm}
\nonumber \\
- \left( g_{\mu_1 \mu_3} p_{\mu_2} + g_{\mu_2 \mu_3} p_{\mu_1} \right)
\; \left[ A(0, p^2; p^2) - B(0, p^2; p^2) \right]
- 2 p_{\mu_1} p_{\mu_2} p_{\mu_3} \; C(p^2, p^2; 0) \; .
\hspace{3mm}
\end{eqnarray}

The one-loop 
contributions to the scalar functions appearing in
(\ref{Gamma_p3=0}) are:
\begin{eqnarray}
\label{AC_p3=0_xi}
\left[ A^{(1,\xi)}(p^2, p^2; 0) + p^2 \; C^{(1,\xi)}(p^2, p^2; 0) \right]
\hspace{86mm}
\nonumber \\
= - \frac{g^2 \; \eta}{(4\pi)^{n/2}} \; 
\frac{C_A \; \kappa}{32 (n\!-\!1)}\; 
\left\{ 8(3n\!-\!4) \!+\! 12\xi (n\!-\!1)(n^2\!-\!4n\!+2) 
\!-\!\xi^2 (n\!-\!1)(n\!-\!4)(n\!+\!2) \right\} ,
\hspace{4mm}
\end{eqnarray}
\begin{eqnarray}
\label{BA_p3=0_xi}
\left[ A^{(1,\xi)}(0, p^2; p^2) - B^{(1,\xi)}(0, p^2; p^2) \right]
= A^{(1,\xi)}(p^2, p^2; 0)
\hspace{60mm}
\nonumber \\
= \frac{g^2 \; \eta}{(4\pi)^{n/2}} \;
\frac{C_A \; \kappa}{8 (n\!-\!1)}\; 
\left\{ 2 (n^2-7n+4) -2\xi (n-1)(4n-13) + \xi^2 (n-1)(n-4) \right\} ,
\hspace{4mm}
\end{eqnarray}
\begin{equation}
\label{C_p3=0_xi}
C^{(1,\xi)}(p^2, p^2; 0) = 
- \frac{g^2 \; \eta}{(4\pi)^{n/2}} \;
\frac{C_A \; (n-4)\; \kappa}{32 (n-1) p^2} \;
\left\{ 8n + 4\xi (n\!-\!1)(3n\!-\!8) - \xi^2 (n\!-\!1)(n\!-\!2) \right\} ;
\end{equation}
\begin{equation}
\left[ A^{(1,q)}(p^2, p^2; 0) + p^2 \; C^{(1,q)}(p^2, p^2; 0) \right]
= \frac{g^2 \; \eta}{(4\pi)^{n/2}} \; N_f T_R \;
\frac{(n-2)^2}{n-1} \; \kappa ,
\end{equation}
\begin{equation}
\label{AB_p3=0_xi}
\left[ A^{(1,q)}(0, p^2; p^2) - B^{(1,q)}(0, p^2; p^2) \right]
= A^{(1,q)}(p^2, p^2; 0) 
= \frac{g^2 \; \eta}{(4\pi)^{n/2}} \; 2 N_f T_R \; 
\frac{n-2}{n-1} \; \kappa ,
\end{equation}
\begin{equation}
\label{Cp_p3=0_xi}
C^{(1,q)}(p^2, p^2; 0) = \frac{g^2 \; \eta}{(4\pi)^{n/2}} \; N_f T_R \;
\frac{(n-4) (n-2)}{n-1} \;
\frac{\kappa}{p^2} ,
\end{equation}
where $\kappa \equiv \kappa(p^2)$.

We note that, according to eqs.~(\ref{BA_p3=0_xi}) and 
(\ref{AB_p3=0_xi}), the following
relation holds for the zero-momentum case:
\begin{equation}
\label{AAB}
A^{(1)}(p^2, p^2; 0) - A^{(1)}(0, p^2; p^2) 
+ B^{(1)}(0, p^2; p^2) = 0 . 
\end{equation}
This also follows from eq.~(\ref{A-B-relation}). 
The relation (\ref{AAB}) is valid
for arbitrary values of $n$ and $\xi$. 
Using (\ref{AAB}), we can reduce the number of tensor structures
in (\ref{Gamma_p3=0}) from three to two, namely\footnote{This corresponds
to the decomposition used in ref.~\cite{BL}, eq.~(A2).}: 
\begin{eqnarray}
\label{BL-decomp}
\Gamma_{\mu_1 \mu_2 \mu_3}^{(1)}(p, -p, 0)
= \left( 2 g_{\mu_1 \mu_2} p_{\mu_3} - g_{\mu_1 \mu_3} p_{\mu_2}
         - g_{\mu_2 \mu_3} p_{\mu_1} \right) \; A^{(1)}(p^2, p^2; 0)
\nonumber \\
 + 2 p_{\mu_3} \left( p^2 g_{\mu_1 \mu_2} - p_{\mu_1} p_{\mu_2} \right) \;
                                               C^{(1)}(p^2, p^2; 0) ,
\end{eqnarray}
where the results for the scalar functions are given
in eqs.~(\ref{BA_p3=0_xi}), (\ref{C_p3=0_xi}), 
(\ref{AB_p3=0_xi}) and (\ref{Cp_p3=0_xi}).
Note that the first tensor structure on the r.h.s.\ of (\ref{BL-decomp})
coincides with the ``zero-loop'' vertex structure, 
given by (\ref{2eq:tree-level}).

To make the complete comparison of our expressions 
(\ref{AC_p3=0_xi})--(\ref{C_p3=0_xi}) (contributing to the
three-gluon vertex at $p_3=0$, eq.~(\ref{Gamma_p3=0}) ) 
with eq.~(20) of \cite{BF},
we need to renormalize our coefficients (\ref{AC_p3=0_xi}) and
(\ref{BA_p3=0_xi}) by adding the counterterm (\ref{A_CT})
to all $A$ functions involved and putting $n=4$. 
Performing this in the $\overline{\mbox{MS}}$ scheme (i.e. at $R=0$),
we find that our results give the same as eq.~(20) of \cite{BF}.  

We have also compared the renormalized (in the $\overline{\mbox{MS}}$
scheme) version of eq.~(\ref{BL-decomp}) with the one-loop results
presented in ref.~\cite{BL}, eq.~(A10). According to 
eq.~(\ref{BL-decomp}) there 
should be the following correspondence between the
functions $T_1$ and $T_2$ used in \cite{BL} and $A^{(1)}$ and $C^{(1)}$:
\begin{equation}
T_1(p^2)  \leftrightarrow A^{(1)}(p^2, p^2; 0) , \hspace{7mm}
T_2(p^2)  \leftrightarrow -2 p^2 C^{(1)}(p^2, p^2; 0) .
\end{equation}
Renormalizing $A^{(1)}(p^2, p^2; 0)$ (given by the sum of
eqs.~(\ref{BA_p3=0_xi}) 
and (\ref{AB_p3=0_xi})) and putting
$n=4$, we arrive at the same result for $T_1$ as given
in eq.~(A10) of ref.~\cite{BL}\footnote{The gauge parameter 
used in \cite{BL} corresponds to our $1-\xi$. To avoid confusion,
we shall call their parameter $\xi_{BL}=1-\xi$. Their constant
$T$ corresponds to our $N_f T_R$. The results presented in
\cite{BL} are taken at $p^2=-\mu^2$, where $\mu^2$ corresponds
to our $\mu_{DR}^2$ (see the discussion in section~2).
Thus, putting $p^2=-\mu^2$ formally corresponds to omitting the terms
containing $\ln(-p^2)$ (which appear due to the expansion of 
$\kappa(p^2)$ in $\varepsilon$) in the renormalized expressions.}.
The result for $T_2$ is finite as $n\to 4$ and should
correspond to the sum of our eqs.~(\ref{C_p3=0_xi}) and 
(\ref{Cp_p3=0_xi}) taken at $n=4$.
However, our result for $T_2$ is different
from the one given in eq.~(A10) of \cite{BL}\footnote{Their result
for $T_2$ is proportional to
$\left[ \left( - {\textstyle{5\over3}} + \xi_{BL} \right) C_A 
       + {\textstyle{4\over3}} T \right]$,
whereas our expressions yield
$\left[ \left( -{\textstyle{37\over24}} + 
{\textstyle{3\over4}} \xi_{BL} 
+ {\textstyle{1\over8}} \xi_{BL}^2 \right) C_A 
       + {\textstyle{4\over3}} T \right]$.
So, the quark contribution is the same while the sum of gluon and
ghost contributions to $T_2$ is different.
Note, that in the Feynman gauge ($\xi_{BL}=1$) our results coincide 
and yield 
$\left[ -{\textstyle{2\over3}} C_A + {\textstyle{4\over3}} T \right]$.
Thus, the disagreement does not influence the one-loop part of
the two-loop result presented in the Feynman gauge, eq.~(B4) of 
\cite{BL}.}. 


\subsection{Two external momenta squared are zero, $p_1^2=p_2^2=0$}

In this case, we substitute
\begin{equation}
p_1^2=p_2^2=0 , \;\; p_3^2 \equiv p^2 , 
\;\; (p_1 p_2)= \textstyle{1\over2} p^2 , \;\;
(p_1 p_3) = (p_2 p_3) = -\textstyle{1\over2} p^2 ;
\end{equation}
\begin{equation}
\label{J111_2}
\left. \frac{}{} J(1,1,1) \right|_{p_1^2=p_2^2=0}
= \mbox{i} \; \pi^{2-\varepsilon}\; \eta \; \varphi(0,0,p^2)
= - \mbox{i} \; \pi^{2-\varepsilon}\; 
\eta 
\; \frac{1}{\varepsilon^2} \; (-p^2)^{-1-\varepsilon} ,
\end{equation}
\begin{equation}
\left. \frac{}{} J(1,0,1) \right|_{p_1^2=p_2^2=0} =
\left. \frac{}{} J(0,1,1) \right|_{p_1^2=p_2^2=0} = 0 ,
\end{equation}
while the result for $J(1,1,0)$ remains unchanged, eq.~(\ref{J(1,1,0)}).
Note that now, when two external lines are on shell, the infrared
singularity in $J(1,1,1)$ is stronger and gives $1/\varepsilon^2$, 
eq.~(\ref{J111_2}).

Again, it is enough to make the above substitutions to get the
result in the Feynman gauge ($\xi=0$), and in the singular gauge
($\xi=-4/(n-4)$), but the situation is more tricky for arbitrary $\xi$
since we have $p_1^2$ and $p_2^2$ in the denominators
of the scalar functions. To solve this problem, we need to
consider the expansion of $J(1,1,1)$ in $p_1^2$ and $p_2^2$.

Two independent ways were used to get the results for the scalar
functions in this limit: \\
(i) \ We take the expressions for one of the momenta squared
equal to zero (see Section~4.1), and put the second momentum
squared equal to zero. In the corresponding expressions 
(see Appendix~E), all  
$p_i^2$ occurring in the denominators are always accompanied
by the corresponding $\kappa_i\equiv\kappa(p_i^2)$ in the 
numerator, which should be put equal to zero when $p_i^2$ vanishes. \\
(ii) \ First, we put $J(1,0,1)=J(0,1,1)=0$. Since the resulting expressions
did not have singularities for $p_1^2=p_2^2$, the next step was
to put $p_1^2=p_2^2\equiv p_0^2$. Then, the integral $J(1,1,1)$
was expanded in $p_0^2/p^2$ keeping the terms up to $(p_0^2/p^2)^2$.
And finally, the limit $p_0^2 \to 0$ was taken. 
The following formula was used to expand $J(1,1,1)$:
\begin{equation}
\left. \frac{}{} J(1,1,1) \right|_{p_1^2=p_2^2\equiv p_0^2}
= - \frac{\mbox{i} \; \pi^{2-\varepsilon}}{(-p^2)^{1+\varepsilon}}\;
\eta \; \frac{1}{\varepsilon^2} \;
_2F_1\left( \left. 
\begin{array}{c} 1, \; \textstyle{1\over2}+\varepsilon \\ 
1+2\varepsilon \end{array} 
\right|
\frac{4 p_0^2}{p^2} \right) ,
\end{equation}
\begin{equation}
_2F_1\left( \left.
\begin{array}{c} 1, \; {\textstyle{1\over2}}+\varepsilon \\ 
1+2\varepsilon \end{array}
\right|
\frac{4 p_0^2}{p^2} \right)
= \sum_{j=0}^{\infty} \left( \frac{4 p_0^2}{p^2} \right)^j
\frac{({\textstyle{1\over2}} +\varepsilon)_j}{(1+2\varepsilon)_j} 
= 1 + 2 \; \frac{p_0^2}{p^2} + 4 \; \frac{n-7}{n-6}
\left(\frac{p_0^2}{p^2}\right)^2 + \ldots 
\end{equation}

The results obtained in these two ways coincide, and
the expressions obtained for the scalar functions (\ref{BC-ggg}) are
presented below. 
Due to the symmetry properties, the functions $A, C$ and $F$ of 
the arguments $(0,p^2,0)$ are equal to 
the corresponding functions of the  arguments $(p^2,0,0)$,
while the $B$ function with these arguments permuted  
changes sign.
The $H$ function is the same for all permutations.

The resulting one-loop contributions (without quark loops) to the scalar 
functions (\ref{BC-ggg}) in arbitrary gauge are: 
\begin{equation}
\label{A(1,xi)(0,0;p^2)}
A^{(1,\xi)}(0,0;p^2)= \frac{g^2 \; \eta}{(4\pi)^{n/2}} \; C_A \;
\frac{1}{32(n-4)} \; \kappa
\left\{ 48 - 8\xi (n-3)(n-6) + \xi^2 (n-4)^2 \right\} ,
\end{equation}
\begin{eqnarray}
A^{(1,\xi)}(p^2,0;0)= - \frac{g^2 \; \eta}{(4\pi)^{n/2}} \; C_A \;
\frac{1}{64(n-1)(n-4)} \; \kappa
\hspace{55mm}
\nonumber \\
\times
\left\{ 16 (2n^2-13n+8) + 4\xi (n-1)(2n-9)(5n-16) 
       - 5\xi^2 (n-1)(n-4)^2 \right\} ,
\end{eqnarray}
\begin{equation}
B^{(1,\xi)}(0,0;p^2)=0,
\end{equation}
\begin{eqnarray}
B^{(1,\xi)}(p^2,0;0)= - \frac{g^2 \; \eta}{(4\pi)^{n/2}} \; C_A \;
\frac{1}{64(n-1)(n-4)} \; \kappa
\hspace{55mm}
\nonumber \\
\times
\left\{ 16 (4n^2-21n+14) + 4\xi (n-1)(10n^2-79n+152)
       - 5\xi^2 (n-1)(n-4)^2 \right\} ,
\end{eqnarray}
\begin{eqnarray}
C^{(1,\xi)}(0,0;p^2)= \frac{g^2 \; \eta}{(4\pi)^{n/2}} \; C_A \;
\frac{1}{4(n-4)(n-6) p^2} \; \kappa
\hspace{45mm}
\nonumber \\
\times
\left\{ 12(n-6) + 2\xi (n^2-11n+36) - \xi^2 (n-3)(n-8) \right\} ,
\end{eqnarray}
\begin{eqnarray}
C^{(1,\xi)}(p^2,0;0)= - \frac{g^2 \; \eta}{(4\pi)^{n/2}} \; C_A \;
\frac{1}{32(n-1)(n-4) p^2} \; \kappa
\hspace{45mm}
\nonumber \\
\times
\left\{ 16 n(n-4) + 4\xi (n-1)(6n^2-41n+72) - 3\xi^2 (n-1)(n-4)^2 
\right\} ,
\end{eqnarray}
\begin{eqnarray}
F^{(1,\xi)}(0,0;p^2)=  \frac{g^2 \; \eta}{(4\pi)^{n/2}} \; C_A \;
\frac{1}{2(n-1)(n-4)(n-6) (p^2)^2} \; \kappa
\hspace{29mm}
\nonumber \\
\times
\left\{ 4(n-3)(n-6)(n-13) - 2\xi (n-1)(3n^2-20n+18) 
\right. 
\hspace{25mm}
\nonumber \\
\left. 
+ \xi^2 (n-1)(n^3-16n^2+74n-78)
- \xi^3 (n-1)(n-3)(n-8) \right\} ,
\end{eqnarray}
\begin{eqnarray}
F^{(1,\xi)}(p^2,0;0)= - \frac{g^2 \; \eta}{(4\pi)^{n/2}} \; C_A \;
\frac{1}{16 (n-1)(n-4)(n-6) (p^2)^2} \; \kappa
\hspace{30mm}
\nonumber \\
\times
\left\{ 16 (n-6)^2 (4n-7) + 4\xi (n-1)(n-6)(10n^2-87n+152) 
\right.
\hspace{25mm}
\nonumber \\
\left.
+ \xi^2 (n-1)(5n^3-94n^2+504n-624)
- 8 \xi^3 (n-1)(n-3)(n-8)
\right\} ,
\end{eqnarray}
\begin{eqnarray}
\label{H(1,xi)(0,0;p^2)}
H^{(1,\xi)}(0,0,p^2)= - \frac{g^2 \; \eta}{(4\pi)^{n/2}} \; C_A \;
\frac{1}{16 (n-1)(n-4)(n-6) p^2} \; \kappa
\hspace{37mm}
\nonumber \\
\times
\left\{ 16 (n-4)(n-6)(n+5) + 24\xi (n-1)(n-6)(n^2-6n+12)
\right.
\hspace{25mm}
\nonumber \\
\left.
- \xi^2 (n-1)(7n^3-110n^2+532n-696)
+ 4\xi^3 (n-1)(n-3)(n-8)
\right\} .
\end{eqnarray}

The quark loop contributions yield:
\begin{eqnarray}
&&
A^{(1,q)}(0,0;p^2)= B^{(1,q)}(0,0;p^2) = C^{(1,q)}(0,0;p^2) = 0 , \\
&&
A^{(1,q)}(p^2,0;0) = B^{(1,q)}(p^2,0;0) 
= \frac{g^2 \; \eta}{(4\pi)^{n/2}} \; N_f T_R \;
\frac{n-2}{n-1} \; \kappa , \\
&&
C^{(1,q)}(p^2,0;0) = \frac{g^2 \; \eta}{(4\pi)^{n/2}} \; 2 N_f T_R \; 
\frac{n-2}{n-1} \; \frac{\kappa}{p^2} , \\
&&
F^{(1,q)}(0,0;p^2)= \frac{g^2 \; \eta}{(4\pi)^{n/2}} \; 64 N_f T_R \;
\frac{1}{(n-1)(n-2)} \; \frac{\kappa}{(p^2)^2} , \\
&&
F^{(1,q)}(p^2,0;0) =  \frac{g^2 \; \eta}{(4\pi)^{n/2}} \; 4 N_f T_R \;
\frac{n (n-6)}{(n-1)(n-2)} 
\; \frac{\kappa}{(p^2)^2} , \\
\label{H(1,q)(0,0;p^2)}
&&
H^{(1,q)}(0,0,p^2) =  \frac{g^2 \; \eta}{(4\pi)^{n/2}} \; 4 N_f T_R \;
\frac{n^2-3n+8}{(n-1)(n-2)} \; 
\; \frac{\kappa}{p^2} .
\end{eqnarray}

Since in the limit $p_1^2=p_2^2=0$ the scalar functions may depend
on $p_3^2 \equiv p^2$ only, the independent tensor structures
(expressed in terms of $p_1$ and $p_2$) can be chosen antisymmetric with 
respect to the permutation $(p_1, \mu_1) \leftrightarrow (p_2, \mu_2)$.
The three-gluon vertex can in this limit be written as
\begin{eqnarray}
\label{ggg_U_i}
\left. \frac{}{}
\Gamma_{\mu_1 \mu_2 \mu_3}(p_1, p_2, p_3)\right|_{p_1^2=p_2^2=0}
= g_{\mu_1 \mu_2} (p_1-p_2)_{\mu_3} \; U_1(p^2)
+ \left[ g_{\mu_1 \mu_3} {p_1}_{\mu_2} - g_{\mu_2 \mu_3} {p_2}_{\mu_1} 
  \right] \; U_2(p^2)
\hspace{3mm}
\nonumber \\
+\left[ g_{\mu_1 \mu_3} {p_2}_{\mu_2} - g_{\mu_2 \mu_3} {p_1}_{\mu_1}
  \right] \; U_3(p^2)
+ {p_1}_{\mu_1} {p_2}_{\mu_2} (p_1-p_2)_{\mu_3} \; U_4(p^2)
\hspace{17mm}
\nonumber \\
+ {p_1}_{\mu_2} {p_2}_{\mu_1} (p_1-p_2)_{\mu_3} \; U_5(p^2)
+ \left[ {p_1}_{\mu_1} {p_1}_{\mu_2} {p_1}_{\mu_3} 
        -{p_2}_{\mu_1} {p_2}_{\mu_2} {p_2}_{\mu_3} \right] \; U_6(p^2)
\hspace{8mm}
\nonumber \\
+ \left[ {p_1}_{\mu_1} {p_1}_{\mu_2} {p_2}_{\mu_3}
        -{p_2}_{\mu_1} {p_2}_{\mu_2} {p_1}_{\mu_3} \right] \; U_7(p^2) .
\hspace{40mm}
\end{eqnarray}
This decomposition is analogous to eq.~(29) of \cite{BF},
and the functions $U_i$ are proportional to the functions $F_j$
used in \cite{BF}
(we have different numbering, see in Appendix~F).

The following representations of the $U_i$ functions in terms of
scalar functions corresponding to the decomposition
(\ref{BC-ggg}) can be derived in this limit:
\begin{equation}
\label{U_1}
U_1(p^2) = A(0,0;p^2) - \textstyle{1\over2} \; p^2 \; C(0,0;p^2)
          - \textstyle{1\over4} \; (p^2)^2 \; F(0,0;p^2)
          + \textstyle{1\over2} \; p^2 \; H(0,0,p^2) ,
\end{equation}
\begin{equation}
U_2(p^2) = -2 \; A(p^2,0;0) - p^2 \; C(p^2,0;0) ,
\end{equation}
\begin{equation}
U_3(p^2) = -A(p^2,0;0) - B(p^2,0;0) 
           - \textstyle{1\over2} \; p^2 \; C(p^2,0;0)
+ \textstyle{1\over4} \; (p^2)^2 \; F(p^2,0;0) 
+ \textstyle{1\over2} \; p^2 \; H(0,0,p^2) ,
\end{equation}
\begin{equation}
U_4(p^2) = C(p^2,0;0) - \textstyle{1\over2} \; p^2 \; F(p^2,0;0)
\end{equation}
\begin{equation}
U_5(p^2) = 2 \; C(p^2,0;0) + C(0,0;p^2) 
+ \textstyle{1\over2} \; p^2 \; F(0,0;p^2) - H(0,0,p^2) ,
\end{equation}
\begin{equation}
U_6(p^2) = 2 \; C(p^2,0;0) ,
\end{equation}
\begin{equation}
\label{U_7}
U_7(p^2) = - C(p^2,0;0) + \textstyle{1\over2} \; p^2 \; F(p^2,0;0)
+ H(0,0,p^2) .
\end{equation}
The explicit results for the $U_i$ functions are presented in
Appendix~F. The infrared-divergent contributions were successfully
compared with eq.~(30) of \cite{BF}, where the corresponding contributions
to $F_j$ are presented.
Note that $U_4, U_5, U_6$ and $U_7$ can be directly compared with \cite{BF},
while the expressions for $U_1, U_2$ and $U_3$
(containing the $A$-function) should be renormalized
by adding the conterterm contribution (\ref{A_CT}) to all $A$ functions
involved.

The result for the three-gluon vertex in the Feynman gauge 
(for $p_1^2=p_2^2=0$) is available in Appendix~B of ref.~\cite{NPS}. 
It is expanded around $n=4$, and the divergent and finite (in $\varepsilon$)
parts are presented. In this limit, our expressions yield the same
as the results of \cite{NPS}. 

\section{Conclusions}
\setcounter{equation}{0}

In this paper, we have presented 
results for the one-loop three-gluon vertex valid for arbitrary values
of the space-time dimension, $n$, and the covariant-gauge parameter,
$\xi$. We have considered the general off-shell case (arbitrary
$p_1^2, p_2^2, p_3^2$, see section~3), as well as all on-shell
cases of interest (section~4). Moreover, having the results
in arbitrary dimension, it was possible to get all on-shell
expressions just by considering the corresponding limits of
the general (off-shell) results. This would be impossible if one
started from the off-shell results expanded around $n=4$,
because in this case the infrared (on-shell) divergencies would appear as
logarithms of vanishing momenta squared.
The only restriction we used in our calculations was that 
in the quark loop contributions the quarks were taken to be
massless\footnote{This restriction does not affect any of the
results for the gluon and ghost loop contributions (the functions
marked $(1,\xi)$) which are indeed the most general ones.}. 

To calculate the vertex, we used the decomposition (\ref{BC-ggg}) 
(adopted from the paper \cite{BC2}) and considered the six scalar 
functions, $A, B, C, S, F$ and $H$, which completely define the 
three-gluon vertex. One of these functions, namely the $S$ function,
was found to be identically zero at the one-loop order\footnote{
In the Feynman gauge and four dimensions, this results was obtained
in \cite{BC2}.}, 
see eqs.~(\ref{S(1,xi)}) and (\ref{S(1,q)}).
We have also checked that $S=0$ when massive quarks are
considered. It is not clear whether it vanishes also 
at the two-loop level.
For the five remaining functions, $A, B, C, F$ and $H$, the general
off-shell
results are given in eqs.~(\ref{A(1,0)})--(\ref{H(1,0)}) (Feynman gauge),
eqs.~(\ref{A(1,xi)})--(\ref{H(1,xi)}) (arbitrary gauge) and
eqs.~(\ref{A(1,q)})--(\ref{H(1,q)}) (quark loop contributions).
They involve only one non-trivial function, $\varphi(p_1^2,p_2^2,p_3^2)$
(see eq.~(\ref{J(1,1,1)}) and Appendix~B), which is related to the
scalar one-loop triangle diagram. For special cases, we have 
successfully compared our results with those from the papers 
\cite{CG,BC2} (for details, see section~3).

Starting from general expressions and putting some external
momenta squared equal to zero, we considered the on-shell cases
$p_3^2=0$ (section~4.1), $p_3=0$ (section~4.2) and $p_1^2=p_2^2=0$
(section 4.3).
For all these cases, the results in arbitrary gauge and dimension
were presented, see section~4 and Appendices~E and F.
For special cases, our results have been compared with those 
presented in refs.~\cite{BF,NPS}. Thus, we can see that  
Table~1 from the Introduction is completely filled in!
Moreover, all results are valid for an arbitrary value of
the space-time dimension. Thus, the only thing which at the one-loop level 
is left for the future is to allow for non-zero quark masses
in the quark loops.

Furthermore, we have obtained general results
for the ghost-gluon vertex (\ref{BC-ghg}), 
see eqs.~(\ref{a(1,xi)})--(\ref{e(1,xi)}) in Appendix~D.  
Employing these results, together with two-point contributions
listed in Appendix~C, we have checked that the Ward--Slavnov--Taylor 
identity (\ref{WST}) for the three-gluon vertex is {\em exactly}
(i.e.\ for arbitrary $n$ and $\xi$) satisfied by the expressions obtained,
as it has to! This was another non-trivial check on the longitudinal
part of the vertex ($A,B,C$ and $S$ functions). 

We note that techniques have recently become available \cite{UD1,UD3} 
to study the off-shell massless vertices at the two-loop level,
at least in the Feynman gauge. Here, the main difficulty are
the integrals with higher powers of irreducible numerators \cite{UD3}.

In the future, the one-loop quark-gluon vertex can also be 
considered in a similar way. For the case of massless quarks,
one can use the same approach as in this paper. For massive
quarks, one should study in more detail what the corresponding
scalar integrals in arbitrary dimension are\footnote{This is also the 
reason why we did not consider massive quark loops in this paper.}.
We note that some results for the quark-gluon vertex (and also
for the QED vertex which formally corresponds to one of the two diagrams 
contributing to the quark-gluon vertex) can be found e.g.\ in 
\cite{qqg,BL}.    
 
\vspace{5mm}

{\bf Acknowledgements.} The authors are indebted to J.S.~Ball and
T.-W.~Chiu for useful communications and, in particular, for confirming 
two minor misprints in \cite{BC2} mentioned in section~3.1 and
Appendix~D. 
O.T. is grateful to NORDITA for supporting his visit to Bergen
where this work was initiated, and to the University of Bergen
for hospitality. 
A.D.'s and P.O.'s research was supported by the Research Council of 
Norway.

\newpage

\appendix
\section*{Appendix A: Decomposition of three-gluon vertex}
\setcounter{equation}{0}
\renewcommand{\thesection}{A}

If we express $p_3$ in terms of the two other momenta,
$p_3=-p_1-p_2$, we get the following decompostion of
the three-gluon vertex (\ref{ggg}):
\begin{eqnarray}
\Gamma_{\mu_1 \mu_2 \mu_3}(p_1, p_2, p_3)
&=&  g_{\mu_1 \mu_2} {p_1}_{\mu_3} Z_{001}
 + g_{\mu_1 \mu_3} {p_1}_{\mu_2} Z_{010}
 + g_{\mu_2 \mu_3} {p_1}_{\mu_1} Z_{100}
\hspace{17mm}
\nonumber \\
&& + g_{\mu_1 \mu_2} {p_2}_{\mu_3} Z_{002}
 + g_{\mu_1 \mu_3} {p_2}_{\mu_2} Z_{020}
 + g_{\mu_2 \mu_3} {p_2}_{\mu_1} Z_{200}
\hspace{17mm}
\nonumber \\
&& + {p_1}_{\mu_1} {p_1}_{\mu_2} {p_1}_{\mu_3} Z_{111}
+ {p_2}_{\mu_1} {p_2}_{\mu_2} {p_2}_{\mu_3} Z_{222}
\hspace{34mm}
\nonumber \\
&& + {p_1}_{\mu_1} {p_1}_{\mu_2} {p_2}_{\mu_3} Z_{112}
+ {p_1}_{\mu_1} {p_2}_{\mu_2} {p_1}_{\mu_3} Z_{121}
+ {p_2}_{\mu_1} {p_1}_{\mu_2} {p_1}_{\mu_3} Z_{211}
\nonumber \\
&& + {p_1}_{\mu_1} {p_2}_{\mu_2} {p_2}_{\mu_3} Z_{122}
+ {p_2}_{\mu_1} {p_1}_{\mu_2} {p_2}_{\mu_3} Z_{212}
+ {p_2}_{\mu_1} {p_2}_{\mu_2} {p_1}_{\mu_3} Z_{221} ,
\end{eqnarray}
where $Z_{jkl}$ are scalar functions depending on $p_1^2, p_2^2$ and
$p_3^2$.

Comparison with the decomposition (\ref{BC-ggg}) used in \cite{BC2}
gives the following
representations of $Z$'s in terms of 
the functions (\ref{BC-ggg}) used by Ball and Chiu \cite{BC2}:
\begin{equation}
Z_{001} = A(p_1^2, p_2^2; p_3^2) - (p_1 p_2) C(p_1^2, p_2^2; p_3^2)
         + B(p_1^2, p_2^2; p_3^2)
         + (p_1 p_2)(p_2 p_3) F(p_1^2, p_2^2; p_3^2) - (p_2 p_3) H ,
\end{equation}
\begin{equation}
Z_{002} = - A(p_1^2, p_2^2; p_3^2) \!+ (p_1 p_2) C(p_1^2, p_2^2; p_3^2)
         \!+ B(p_1^2, p_2^2; p_3^2)
         \!- (p_1 p_2)(p_1 p_3) F(p_1^2, p_2^2; p_3^2) \!+(p_1 p_3) H ,
\end{equation}
\begin{equation}
Z_{100} = A(p_2^2, p_3^2; p_1^2) - (p_2 p_3) C(p_2^2, p_3^2; p_1^2)
         - B(p_2^2, p_3^2; p_1^2)
         + (p_1 p_2)(p_2 p_3) F(p_2^2, p_3^2; p_1^2) - (p_1 p_2) H ,
\end{equation}
\begin{equation}
Z_{200} = 2 A(p_2^2, p_3^2; p_1^2) - 2 (p_2 p_3) C(p_2^2, p_3^2; p_1^2)
         - p_1^2 (p_2 p_3) F(p_2^2, p_3^2; p_1^2) + p_1^2 H ,
\end{equation}
\begin{equation}
Z_{010} = - 2 A(p_3^2, p_1^2; p_2^2) + 2 (p_1 p_3) C(p_3^2, p_1^2; p_2^2)
         + p_2^2 (p_1 p_3) F(p_3^2, p_1^2; p_2^2) - p_2^2 H ,
\end{equation}
\begin{equation}
Z_{020} = - A(p_3^2, p_1^2; p_2^2) \!+ (p_1 p_3) C(p_3^2, p_1^2; p_2^2)
         \!- B(p_3^2, p_1^2; p_2^2)
         \!- (p_1 p_2)(p_1 p_3) F(p_3^2, p_1^2; p_2^2) \!+ (p_1 p_2) H ,
\end{equation}
\begin{equation}
Z_{111} = 2 C(p_3^2, p_1^2; p_2^2) + p_2^2 F(p_3^2, p_1^2; p_2^2),
\end{equation}
\begin{equation}
Z_{222} = -2 C(p_2^2, p_3^2; p_1^2) - p_1^2 F(p_2^2, p_3^2; p_1^2),
\end{equation}
\begin{equation}
Z_{112} = - C(p_2^2, p_3^2; p_1^2) + (p_1 p_2) F(p_2^2, p_3^2; p_1^2)
          + H - S ,
\end{equation}
\begin{equation}
Z_{121} = C(p_3^2, p_1^2; p_2^2) - (p_1 p_2) F(p_3^2, p_1^2; p_2^2) ,
\end{equation}
\begin{equation}
Z_{122} = - C(p_2^2, p_3^2; p_1^2) + (p_1 p_2) F(p_2^2, p_3^2; p_1^2) ,
\end{equation}
\begin{equation}
Z_{211} = C(p_1^2, p_2^2; p_3^2) + 2 C(p_3^2, p_1^2; p_2^2)
          - (p_2 p_3) F(p_1^2, p_2^2; p_3^2)
          + p_2^2 F(p_3^2, p_1^2; p_2^2) - H - S ,
\end{equation}
\begin{equation}
Z_{212} = -  C(p_1^2, p_2^2; p_3^2) - 2 C(p_2^2, p_3^2; p_1^2)
          + (p_1 p_3) F(p_1^2, p_2^2; p_3^2)
          - p_1^2 F(p_2^2, p_3^2; p_1^2) + H - S ,
\end{equation}
\begin{equation}
Z_{221} = C(p_3^2, p_1^2; p_2^2) - (p_1 p_2) F(p_3^2, p_1^2; p_2^2)
          - H - S ,
\end{equation}
where $H\equiv H(p_1^2, p_2^2, p_3^2)$ and 
$S\equiv S(p_1^2, p_2^2, p_3^2)$.

Solving these equations we get the following results for
the scalar functions (\ref{BC-ggg}),
including those with permuted arguments, in terms of $Z$'s:
\begin{equation}
S(p_1^2, p_2^2, p_3^2) =
{\textstyle{1\over2}} 
\left\{ - Z_{112} + Z_{121} + Z_{122} - Z_{221} \right\} ,
\end{equation}
\begin{equation}
H(p_1^2, p_2^2, p_3^2) =
{\textstyle{1\over2}} 
\left\{ Z_{112} + Z_{121} - Z_{122} - Z_{221} \right\} ,
\end{equation}
\begin{equation}
A(p_1^2, p_2^2; p_3^2) =
{\textstyle{1\over2}} \left\{ \frac{}{}
(p_1 p_2) \left[ - Z_{111} + Z_{222} + Z_{211} - Z_{212} \right]
+ Z_{001} - Z_{002} - (p_1^2 + p_2^2) H  \right\} ,
\end{equation}
\begin{equation}
A(p_2^2, p_3^2; p_1^2) =
{\textstyle{1\over2}} 
\left\{ - (p_2 p_3) Z_{222} + Z_{200} - p_1^2 H \right\} ,
\end{equation}
\begin{equation}
A(p_3^2, p_1^2; p_2^2) =
{\textstyle{1\over2}} 
\left\{ (p_1 p_3) Z_{111} + Z_{010} - p_2^2 H \right\} .
\end{equation}
\begin{eqnarray}
B(p_1^2, p_2^2; p_3^2) =
{\textstyle{1\over2}} \left\{
p_1^2 \left[ Z_{112} - Z_{122} \right]
+ p_2^2 \left[ - Z_{121} + Z_{221} \right]
\right.
\hspace{4cm}
\nonumber \\
\left.
- (p_1 p_2) \left[ Z_{111} + Z_{222} - Z_{211} - Z_{212} \right]
+ Z_{001} + Z_{002} + p_3^2 S \right\} ,
\end{eqnarray}
\begin{equation}
B(p_2^2, p_3^2; p_1^2) =
{\textstyle{1\over2}} 
\left\{-(p_2 p_3)(Z_{222}-2 Z_{122}) + Z_{200} - 2 Z_{100}
                   + (p_2^2 - p_3^2) H \right\} ,
\end{equation}
\begin{equation}
B(p_3^2, p_1^2; p_2^2) =
{\textstyle{1\over2}} 
\left\{-(p_1 p_3)(Z_{111}-2 Z_{121}) + Z_{010} - 2 Z_{020}
                   + (p_3^2 - p_1^2) H \right\} ,
\end{equation}
\begin{eqnarray}
C(p_1^2, p_2^2; p_3^2) =
\frac{1}{p_1^2 - p_2^2}
\left\{ (p_1 p_3) \left[ Z_{111} - Z_{211} - Z_{121} + Z_{221} \right]
\right.
\hspace{5mm}
\nonumber \\
\left.
+ (p_2 p_3) \left[ Z_{222} + Z_{112} - Z_{122} - Z_{212} \right]
\right\} ,
\end{eqnarray}
\begin{equation}
C(p_2^2, p_3^2; p_1^2) =
\frac{1}{p_2^2 - p_3^2}
\left\{ (p_1 p_2) Z_{222} + p_1^2 Z_{122} \right\}  ,
\end{equation}
\begin{equation}
C(p_3^2, p_1^2; p_2^2) =
\frac{1}{p_3^2 - p_1^2}
\left\{ (p_1 p_2) Z_{111} + p_2^2 Z_{121} \right\}  ,
\end{equation}
\begin{equation}
F(p_1^2, p_2^2; p_3^2) =
\frac{1}{p_1^2 - p_2^2}
\left\{ Z_{111} + Z_{222} + Z_{112} - Z_{121} - Z_{211}
       - Z_{122} - Z_{212} + Z_{221} \right\} ,
\end{equation}
\begin{equation}
F(p_2^2, p_3^2; p_1^2) =
\frac{1}{p_2^2 - p_3^2}
\left\{ Z_{222} - 2 Z_{122} \right\}  ,
\end{equation}
\begin{equation}
F(p_3^2, p_1^2; p_2^2) =
\frac{1}{p_3^2 - p_1^2}
\left\{ Z_{111} - 2 Z_{121} \right\}  .
\end{equation}

\appendix
\section*{Appendix B: Scalar integrals}
\setcounter{equation}{0}
\renewcommand{\thesection}{B}

As mentioned in section~2,
the results for the scalar functions occurring in eqs.~(\ref{BC-ggg})
and (\ref{BC-ghg}) can be represented in terms of the following
Feynman integrals, corresponding to a scalar one-loop triangle
diagram:
\begin{equation}
\label{defJ_2}
J (\nu_1  ,\nu_2  ,\nu_3) \equiv \int
 \frac{\mbox{d}^n q}{ ((p_2 -q )^2)^{\nu_1}  ((p_1 +q )^2)^{\nu_2}
      (q^2)^{\nu_3} } ,
\end{equation}
where $n=4-2\varepsilon$ is the space-time dimension.

When we perform calculations in the Feynman gauge and express the
scalar numerators in terms of the denominators, the powers
of the denominators, $\nu_i$, can be one, zero, or
even negative. In an arbitrary gauge, the integrals
may have powers of $\nu_i$ equal to two, due to the presence of
$(p^2)^{-2}$ in the $\xi$ term of the gluon
propagator, eq.~(\ref{gl_prop}). Nevertheless, by using the
integration-by-parts technique \cite{ibp} these integrals
can be reduced to those with $\nu$'s equal to one or zero.
The corresponding algorithm for the integrals (\ref{defJ_2}) is 
described in detail in \cite{JPA}\footnote{The main formula to be used
is eq.~(3.4) of \cite{JPA}. Some explicit results for the integrals
with $\nu_i=2$ are also presented in \cite{JPA}.}.

Then, if two or three $\nu$'s are non-positive 
integers, the dimensionally-regularized integral (\ref{defJ_2})
vanishes \cite{dimreg}, since it corresponds to a massless tadpole.
When one of the $\nu$'s is zero, the integral
(\ref{defJ_2}) corresponds to a two-point function.
Therefore, it is proportional
to a power of the external momentum squared times some $\Gamma$ 
functions with arguments involving $n$ and $\nu$'s.
We mainly need the result for two remaining $\nu$'s equal to one
(e.g. $J(1,1,0)$)
which is given by eqs.~(\ref{kappa})--(\ref{J(1,1,0)}),
in which case this $\Gamma$ factor is nothing but $\eta$, eq.~(\ref{eta}).

Then, the integrals with one negative $\nu$ can easily be 
reduced to integrals with the corresponding $\nu$ equal to zero,
for example:
\begin{eqnarray}
J(1,1,-1) &=& - (p_1 p_2) \; J(1,1,0) ,  \\
J(1,1,-2) &=& \frac{1}{n-1} \left[ n (p_1 p_2)^2 - p_1^2 p_2^2 \right] \;
            J(1,1,0), \\
J(1,1,-3) &=& -\frac{1}{n-1} 
\left[ (n+2) (p_1 p_2)^2 - 3 p_1^2 p_2^2 \right] \; (p_1 p_2) \; J(1,1,0) .
\end{eqnarray}

Thus, the only non-trivial function which occurs in our 
calculations
is $\varphi$ which is related to the triangle integral $J(1,1,1)$ 
(cf. (\ref{J(1,1,1)})) via
\begin{equation}
\label{J(1,1,1)_2}
J(1,1,1) = {\mbox{i}} \pi^{n/2} \;
\eta \;
\varphi(p_1^2,p_2^2,p_3^2) ,
\end{equation}
where $\eta$ is defined by eq.~(\ref{eta}).

In fact, the general results (i.e.\ for arbitrary $n$, $\nu_i$ and $p_i^2$)
for the integrals (\ref{defJ_2}) are available \cite{BD}. They can
be represented in terms of Appell's hypergeometric function of
two variables, $F_4$. As dimensionless variables, one can use
\begin{equation}
\label{xy}
 x \equiv \frac{p_1^2}{p_3^2} \hspace{0.5cm}\mbox{and}
\hspace{0.5cm} y \equiv \frac{p_2^2}{p_3^2} \; .
\end{equation}
When $\nu_1=\nu_2=\nu_3=1$, all the
$F_4$ functions can be reduced to $_2F_1$ Gauss hypergeometric
functions of more complicated arguments, 
by using reduction formulae for $F_4$ functions
(see e.g. in \cite{Erdelyi}). One can also derive a one-dimensional
integral representation (see eq.~(26) of \cite{UD3})
which is valid for arbitrary $\varepsilon=(4-n)/2$ and therefore 
for arbitrary $n$,
\begin{equation}
\label{int-dr}
J(1,1,1) 
= -\frac{\mbox{i}\; \pi^{2-\varepsilon} \; \eta}{(-p_3^2)^{1+\varepsilon}} \;
                 \frac{1}{\varepsilon} \int\limits_0^1
\frac{\mbox{d}\sigma\;\sigma^{-\varepsilon} 
                 \left( (y\sigma)^{-\varepsilon} 
                - (x/\sigma)^{-\varepsilon} \right)}
      {\left( y \sigma^2 + (1-x-y)\sigma + x \right)^{1-\varepsilon}} ,
\end{equation}
with $x$ and $y$ defined by eq.~(\ref{xy}).

Another way to get the result for arbitrary $n$ is to use
the connection \cite{DT2} between massless triangle integrals and the
two-loop massive vacuum integrals, $I(\nu_1,\nu_2,\nu_3)$. 
In particular, according to 
eq.~(40) of \cite{DT2},  for arbitrary $n$ the integral $J(1,1,1)$ 
can be related (up to trivial leftovers) to the integral $I(1,1,1)$,
in such a way that one of the integrals is taken in $4-2\varepsilon$ 
dimensions, and the other in $4+2\varepsilon$ dimensions. 
Using this connection,
and also known results for $I(1,1,1)$ in arbitrary dimension
\cite{FJJ,Scharf,DT1}, one can reproduce the result for $J(1,1,1)$.
It can be written in terms of $_2F_1$ functions (see e.g. eq.~(4.12)
of \cite{DT1}).   

The result for $J(1,1,1)$ in four dimensions is well-known 
\cite{'tHV-79,BC2} (see also in \cite{JPA,UD1}) and can be presented in 
terms of dilogarithms,
\begin{equation}
\label{phi_n=4}
\left. \!\! \frac{}{} \varphi(p_1^2,p_2^2,p_3^2) \right|_{n=4}\!
= \frac{1}{p_3^2 \lambda} \left\{ \frac{}{}
2 \left[ \mbox{Li}_{2}\left(-\rho x\right) 
\!+\! \mbox{Li}_{2}\left(-\rho y\right) \right]
+ \ln\frac{y}{x}  \ln{\frac{1\!+\!\rho y}{1\!+\!\rho x}}
+ \ln(\rho x) \ln(\rho y) + \frac{\pi^2}{3}
\right\} ,
\end{equation}
where
\begin{equation}
\label{lambda}
\lambda(x,y) \equiv \sqrt{(1-x-y)^2 - 4 x y} \; \; \; ,
\; \; \; \rho(x,y) \equiv \frac{2}{1-x-y+\lambda} ,
\end{equation}
or, in terms of the Clausen function, $\mbox{Cl}_2$ 
(see e.g. eq.~(19) of \cite{DT2}; 
similar representations are also given in \cite{FJJ,Lu}). 
In particular, in the symmetric case (see section~3.4)
\begin{equation}
\label{Cl2(pi/3)}
\left. 
\frac{}{} J(1,1,1) \right|_{\begin{array}{c}
{}_{p_1^2=p_2^2=p_3^2\equiv p^2} \\ {}^{n=4 \hspace*{10mm}} \end{array}}
= \mbox{i} \pi^2 \; \left. \varphi(p^2,p^2,p^2) \right|_{n=4}
= \frac{\mbox{i} \pi^2}{p^2} \; \frac{4}{\sqrt{3}} \;
\mbox{Cl}_2\left(\frac{\pi}{3}\right) ,
\end{equation}
producing the same constant as the one (denoted as $I$) used in \cite{CG}.
The transcendental number $\mbox{Cl}_2\left(\frac{\pi}{3}\right)
= 1.0149417...$ corresponds to the maximum of Clausen's integral
and appears frequently in two-loop calculations with masses
(see e.g. \cite{Cl,FJJ} ), the connection with the massless triangle
diagrams being clear from \cite{DT2}.

If one is interested in expanding in $\varepsilon$ and calculating
the integral $J(1,1,1)$ up to order $\varepsilon$, one can use
eq.~(29) of \cite{UD3}, or eqs.~(16) and (20) of \cite{DT2}.

In three dimensions, the result for the integral $J(1,1,1)$
is very simple \cite{D'Eramo} and proportional to 
$(p_1^2 p_2^2 p_3^2)^{-1/2}$. To get the result around two
dimensions, eq.~(43) of \cite{DT2} can be used\footnote{This 
is a special case of a more general result presented in \cite{BDK}.}.
The corresponding two-dimensional integral has infrared singularities which
can be regularized by the same $\varepsilon$. However, the only non-trivial
function is the same as in the four-dimensional case. The same 
equation (43) of \cite{DT2} can also be used to get the results 
for higher values of $n$. 

When some of the external momenta squared vanish, the integral
$J(1,1,1)$ (considered in four dimensions) develops infrared (on-shell) 
singularities. The corresponding limits are considered in
section~4.

\appendix
\section*{Appendix C: Two-point functions}
\setcounter{equation}{0}
\renewcommand{\thesection}{C}

To check whether our results are consistent with the
Ward--Slavnov--Taylor identity for the three-gluon vertex (\ref{WST}),
we need expressions for the functions contributing to the gluon
polarization operator and the ghost self energy. 

The corresponding scalar functions, $J(p^2)$ and $G(p^2)$, 
are defined in eqs.~(\ref{gl_po}) and (\ref{gh_se}), respectively.
The lowest-order results are $J^{(0)}=G^{(0)}=1$.
At the one-loop order, the results can be found e.g. in \cite{Muta}.
We present them here for completeness, and also to show the proper
normalization of the functions. One-loop contributions to these
functions are presented in Fig.~2. 

The gluon and ghost loop contributions to the $J$ and $G$ function 
in an arbitrary covariant gauge yield
\begin{equation}
\label{J(1,xi)}
J^{(1,\xi)}(p^2) = - \frac{g^2 \; \eta}{(4\pi)^{n/2}} 
\frac{C_A}{8(n-1)} 
\left\{ 4(3n-2) + 4 \xi (n-1) (2n-7) - \xi^2 (n-1) (n-4) \right\}
 \kappa(p^2) ,
\end{equation}
\begin{equation}
\label{G(1,xi)}
G^{(1,\xi)}(p^2) = \frac{g^2 \; \eta}{(4\pi)^{n/2}} \;
 \frac{C_A}{4}
\left( 2 + \xi (n-3) \right) \; \kappa(p^2)  ,
\end{equation}
while the quark loop contribution to the $J$ function is
\begin{equation}
\label{J(1,q)}
J^{(1,q)}(p^2) = \frac{g^2 \; \eta}{(4\pi)^{n/2}} \;
2 \; N_f T_R \; \frac{n-2}{n-1} \; \kappa(p^2) 
\end{equation}
(there is no quark contribution to the $G$ function). The coefficients
$\eta$, $C_A$ and $T_R$ are defined by eqs.~(\ref{eta}), (\ref{C_A})
and (\ref{T_R}), respectively. The function $\kappa(p^2)$ is given by
eq.~(\ref{kappa}).

The ultraviolet divergences of eqs.~(\ref{J(1,xi)})--(\ref{J(1,q)})
are given by
\begin{equation}
J^{(1,UV)}= \frac{g^2 \; \eta}{(4\pi)^{2-\varepsilon}} 
\left[ -C_A \left( {\textstyle{5\over3}} + {\textstyle{1\over2}} \xi \right)
+{\textstyle{4\over3}} N_f T_R \right] \kappa^{(UV)},
\hspace{4mm}
G^{(1,UV)}= \frac{g^2 \; \eta}{(4\pi)^{2-\varepsilon}} C_A \; 
{\textstyle{1\over4}} (2\!+\!\xi) \kappa^{(UV)},
\end{equation}
where $\kappa^{(UV)} = 1/\varepsilon + \ldots$ is the divergent part of 
$\kappa(p^2)$, eq.~(\ref{kappa}). The corresponding counterterms are
\begin{eqnarray}
J^{(1,CT)} &=& \frac{{\overline{g}}^2}{(4\pi)^2} \;
\left[ 
C_A \left( {\textstyle{5\over3}} + {\textstyle{1\over2}} \xi \right)
- {\textstyle{4\over3}} N_f T_R \right]
\left( \frac{1}{\varepsilon} + R \right)
\nonumber \\
&=& \frac{g^2 \; \eta}{(4\pi)^{2-\varepsilon}} \;
\left[ 
C_A \left( {\textstyle{5\over3}} + {\textstyle{1\over2}} \xi \right)
- {\textstyle{4\over3}} N_f T_R \right]
\left( \frac{1}{\varepsilon} + R \right) + {\cal{O}}(\varepsilon) ,
\nonumber \\
G^{(1,CT)} &=& -\frac{{\overline{g}}^2}{(4\pi)^2} \;
\frac{C_A}{4} (2+\xi)
\left( \frac{1}{\varepsilon} + R \right)
= -\frac{g^2 \; \eta}{(4\pi)^{2-\varepsilon}} \;
\frac{C_A}{4} (2+\xi)
\left( \frac{1}{\varepsilon} + R \right) + {\cal{O}}(\varepsilon) ,
\hspace{6mm}
\end{eqnarray}
where 
${\overline{g}}^2 \equiv g^2 e^{-\gamma \varepsilon} (4\pi)^{\varepsilon}$, and
$R$ is the renormalization-scheme constant chosen in such a way that
$R=0$ in the $\overline{\mbox{MS}}$ scheme (see also section~3.5).

Note that in the Fried--Yennie gauge \cite{FriedYennie}\footnote{This
gauge was also used in refs.~\cite{AbrSol}.}, $\xi=-2$, 
the ghost self-energy is finite as $n\to 4$. Moreover, if one chooses
the $n$-dimensional generalization of this gauge as 
$\xi = -2/(n-3)$ \cite{Diploma}, then the one-loop correction to the ghost 
self-energy vanishes. This is connected with the transversality of the
gluon propagator (\ref{gl_prop}) in the coordinate space (at this
value of $\xi$).

The gluon polarization operator is finite when 
\begin{equation}
\left. \frac{}{} \xi \right|_{n=4} \equiv \xi_0  =
{\textstyle{2\over3}} \left( \frac{4 N_f T_R}{C_A} - 5 \right) .
\end{equation}
For $N_f=6, \; T_R= {\textstyle{1\over2}}$ and $C_A=3$, 
this value is $\xi_0=-{\textstyle{2\over3}}$.

\appendix
\section*{Appendix D: Results for the ghost-gluon vertex}
\setcounter{equation}{0}
\renewcommand{\thesection}{D}

There are two one-loop contributions to the ghost-gluon vertex
which are shown in Fig.~3.
Here, we present the most general results for
the scalar functions contributing to the 
ghost-gluon vertex at the one-loop level.
The definition of the ghost-gluon vertex and the decomposition
in terms of scalar functions is given in eqs.~(\ref{ghg}) and
(\ref{BC-ghg}), respectively.
The lowest-order expression is given by eq.~(\ref{ghg0}).
In one-loop expressions, we use the notation for $\varphi$, 
$\kappa_i$ and $\eta$ which can be found in eqs.~(\ref{J(1,1,1)}),
(\ref{kappa}) and (\ref{eta}), respectively. 
${\cal{K}}$ and ${\cal{Q}}$ denote the symmetric scalar combinations
constructed from the external momenta, eqs.~(\ref{kkk}) and (\ref{qqq}). 

At the one-loop level (in the Feynman gauge) we get
\begin{eqnarray}
\label{ghg(1,0)}
\widetilde{\Gamma}^{(1,0)}_{\mu \mu_3} (p_1, p_2; p_3)
= - \frac{g^2 \; \eta}{(4\pi)^{n/2}} \; C_A \; \frac{1}{4 {\cal{K}}}
\hspace{88mm}
\nonumber \\
\times
\left\{
g_{\mu \mu_3} {\cal{K}}
\left[ \left(2 p_1^2 - p_2^2 \right) \varphi
      + \kappa_1 - 2 \kappa_2 - \kappa_3 \right]
\right.
\hspace{73mm}
\nonumber \\
+ {p_1}_{\mu} {p_1}_{\mu_3}
\left[ p_2^2 (p_1 p_3) \varphi + (p_1 p_2) \kappa_1 + p_2^2 \kappa_2
       + (p_2 p_3) \kappa_3 \right]
\hspace{48mm}
\nonumber \\
- 2 {p_2}_{\mu} {p_2}_{\mu_3}
\left[ p_1^2 (p_2 p_3) \varphi + p_1^2 \kappa_1  + (p_1 p_2) \kappa_2
         + (p_1 p_3) \kappa_3  \right]
\hspace{46mm}
\nonumber \\
+ {p_1}_{\mu} {p_2}_{\mu_3}
\left[ - \left( p_1^2 (p_2 p_3) + p_2^2 (p_1 p_3) \right) \varphi
      + (p_1 p_3) \kappa_1 + (p_2 p_3) \kappa_2 + p_3^2 \kappa_3 \right]
\hspace{20mm}
\nonumber \\
\left.
+ {p_2}_{\mu} {p_1}_{\mu_3}
\left[ \left( p_2^2 (p_1 p_3) + 2 (p_1 p_2)(p_2 p_3) \right) \varphi
      + 3 (p_1 p_2) \kappa_1 + 3 p_2^2 \kappa_2 + 3 (p_2 p_3) \kappa_3 
 \right]
\right\} .
\hspace{5mm}
\end{eqnarray}

For the scalar functions (\ref{BC-ghg}), eq.~(\ref{ghg(1,0)}) yields
\begin{equation}
\label{a(1,0)}
a^{(1,0)}(p_3, p_2, p_1)
= - \frac{g^2 \; \eta}{(4\pi)^{n/2}} \; C_A \; \frac{1}{4} \;
\left\{ (2 p_1^2 - p_2^2) \varphi + \kappa_1 - 2 \kappa_2 - \kappa_3
\right\} ,
\end{equation}
\begin{equation}
\label{b(1,0)}
b^{(1,0)}(p_3, p_2, p_1)
=  \frac{g^2 \; \eta}{(4\pi)^{n/2}} \; C_A \; \frac{1}{2 {\cal{K}} } \;
\left\{ p_1^2 (p_2 p_3) \varphi + p_1^2 \kappa_1 + (p_1 p_2) \kappa_2
+ (p_1 p_3) \kappa_3 \right\} ,
\end{equation}
\begin{eqnarray}
\label{c(1,0)}
c^{(1,0)}(p_3, p_2, p_1)
= - \frac{g^2 \; \eta}{(4\pi)^{n/2}} \; C_A \; \frac{1}{4 {\cal{K}} } \;
\hspace{77mm}
\nonumber \\
\times
\left\{ (p_1^2-p_2^2) (p_1 p_2) \varphi - (p_1^2 - (p_1 p_2)) \kappa_1
+ (p_2^2 - (p_1 p_2)) \kappa_2 + (p_1^2-p_2^2) \kappa_3 \right\} ,
\end{eqnarray}
\begin{eqnarray}
\label{d(1,0)}
d^{(1,0)}(p_3, p_2, p_1)
=  \frac{g^2 \; \eta}{(4\pi)^{n/2}} \; C_A \; \frac{1}{4 {\cal{K}} } \;
\hspace{80mm}
\nonumber \\
\times
\left\{ \left( p_2^2 (p_1 p_3) + 2 (p_1 p_2) (p_2 p_3) \right) \varphi 
+ 3 (p_1 p_2) \kappa_1
+ 3 p_2^2 \kappa_2 + 3 (p_2 p_3) \kappa_3 \right\} ,
\end{eqnarray}
\begin{equation}
\label{e(1,0)}
e^{(1,0)}(p_3, p_2, p_1)
= - \frac{g^2 \; \eta}{(4\pi)^{n/2}} \; C_A \; \frac{1}{4 {\cal{K}} } \;
\left\{ p_3^2 (p_1 p_2) \varphi + (p_1 p_3) \kappa_1 
+ (p_2 p_3) \kappa_2 + p_3^2 \kappa_3 \right\} .
\end{equation} 

In the limit $n\to 4$, the expressions for all scalar functions 
(\ref{a(1,0)})--(\ref{e(1,0)}) have been compared with the corresponding 
results presented in \cite{BC2}, taking into account the erratum
(which affects the results for the $c,d$ and $e$ functions).
The comparison was successful, with the exception of two minor things.
One of them is related to the definition of the renormalization-scheme
constant and was already mentioned before, see footnote in 
section~3.1. The second one is that
in the erratum \cite{BC2} the sign of the term $b(P_1,P_2,P_3)$ 
in the expression for the $d$ function is changed from minus to plus
(see p.~2554 of \cite{BC2}). However,
the comparison is successful if we keep the original sign, 
which is minus\footnote{These misprints were confirmed by the
authors of \cite{BC2}.}.

Furthermore, the results which follow 
from eq.~(\ref{ghg(1,0)}) (contracted with $p_1^{\mu}$)
for two infrared-divergent cases, 
(i) $p_1^2=p_2^2=0$ and (ii)
$p_2^2=p_3^2=0$, have been compared with those presented
in table~B.II of ref.~\cite{NPS}. The latter were obtained in the Feynman
gauge and expanded in the limit $n\to 4$, keeping the finite
(in $\varepsilon=(4-n)/2$) terms. To consider the limit of our expressions,
we used formulae presented in section~4.3. 
We found that our results coincide in this limit with those
from \cite{NPS}. 

Now, we present the results for the scalar functions (\ref{BC-ghg})
in arbitrary covariant gauge:
\begin{eqnarray}
\label{a(1,xi)}
a^{(1,\xi)}(p_3,p_2,p_1) =
- \frac{g^2 \; \eta}{(4\pi)^{n/2}} \; C_A\; \frac{1}{16 {\cal{K}}}
\hspace{87mm}    
\nonumber \\
\times \left\{
 \left[ 2 {\cal{K}} \left( \xi (2 n-5)+2 \right)
        + \xi \left( (n-3) \xi+2 \right) p_2^2 (p_1 p_3)
        - \xi (\xi-2) p_3^2 (p_1 p_2) \right]
\left[ (p_2 p_3) \varphi+\kappa_1 \right]
\right.
\nonumber \\
+ \left[ (\xi-2) \left( 4 {\cal{K}}-\xi p_2^2 p_3^2 \right)
         + \xi \left( (n-3) \xi+2 \right) p_2^2 (p_2 p_3) \right]
\left[ (p_1 p_3) \varphi+\kappa_2 \right]
\hspace{27mm}
\nonumber \\
+ \left[ (\xi-2) \left( (\xi+2) {\cal{K}} - \xi p_3^2 (p_2 p_3) \right)
         + \xi \left( (n-3) \xi+2 \right) (p_2 p_3)^2 \right]
\left[ (p_1 p_2) \varphi+\kappa_3 \right]
\hspace{11mm}
\nonumber \\
\left.
+ \xi (p_2 p_3) \varphi \; 
  \left[ (\xi \!-\!2) \left( {\cal{K}}\!+\! 2 p_3^2 (p_1 p_2) \right)
        + \left( (n\!-\!3) \xi \!+\! 2 \right) 
                \left( {\cal{K}}\!-\! 2 (p_1 p_2) (p_2 p_3) \right) \right]
                                   \right\} , 
\hspace{16mm}
\end{eqnarray}
\begin{eqnarray}
\label{b(1,xi)}
b^{(1,\xi)}(p_3,p_2,p_1) =
- \frac{g^2 \; \eta}{(4\pi)^{n/2}} \; C_A \; \frac{1}{16 {\cal{K}}^2 p_3^2}
\hspace{82mm}
\nonumber \\
\times \! \left\{
\left[ \xi \left( (n-4) \xi+4 \right) (n-3) (p_1 p_2) (p_2 p_3) {\cal{K}}
\right. \right.
\hspace{78mm}
\nonumber \\
+ \left( (n-3) \xi+2 \right) p_1^2 p_3^2 
\left( 2 (\xi-2) {\cal{K}}-\xi (n-1) (p_1 p_2) (p_2 p_3) \right)
\hspace{45mm}
\nonumber \\
\left.
+\xi (\xi-2) p_3^2 
\left( (n-1) p_1^2 p_3^2 (p_1 p_2)+{\cal{K}} (p_1^2-2 (n-3) (p_1 p_2)) \right)
\right]
\left[ (p_2 p_3) \varphi+\kappa_1 \right]
\nonumber \\
+ \left[ -\xi \left( (n\!-\!4) \xi+4 \right) 
(n-3) (p_1 p_2) (p_2 p_3) {\cal{K}}
+ \xi (\xi-2) p_3^2
\left( (n\!-\!1) p_1^2 p_2^2 p_3^2+(n\!-\!2) {\cal{K}} {\cal{Q}} \right)
\right.
\nonumber \\
\left.
+ \left( (n\!-\!3) \xi \!+\! 2 \right) p_3^2 
\left( 
{\cal{K}} (\xi p_2^2 \!+\!(3 \xi\!-\!4) (p_1 p_2)) 
\!-\! \xi (n\!-\!1) (p_1 p_2)^2 (p_2 p_3)
\right)
\right]
\left[ (p_1 p_3) \varphi+\kappa_2 \right]
\nonumber \\
+ \left[ -\xi \left( (n-4) \xi+4 \right) (n-4) (p_1 p_3) (p_2 p_3) {\cal{K}}
\right.
\hspace{73mm}
\nonumber \\
+ \left( (n-3) \xi+2 \right) p_3^2 
\left( \xi (p_2 p_3) ({\cal{K}}-(n-1) (p_1 p_2) (p_1 p_3))
                                     +2 (\xi-2) {\cal{K}} (p_1 p_3) \right)
\hspace{12mm}
\nonumber \\
\left.
+\xi (\xi-2) p_3^2 
\left( (n-1) (p_2 p_3) (p_1 p_3)^2-{\cal{K}} (p_2^2+(p_1 p_3)) \right)
\right]
 \left[ (p_1 p_2) \varphi+\kappa_3 \right]
\nonumber \\
+ \varphi  \left[ 2 \xi \left( (n-4) \xi+4 \right) 
                (n-3) (p_1 p_2) (p_1 p_3) (p_2 p_3) {\cal{K}}
\right.
\hspace{64mm}
\nonumber \\
+\!\left( (n\!-\!3) \xi\!+\!2 \right) p_3^2 
\left( 2 \xi (n\!-\!1) (p_1 p_2)^2 (p_1 p_3) (p_2 p_3)
\!-\!2 {\cal{K}} (p_1 p_3) (2 (\xi\!-\!2) (p_1 p_2)\!+\!\xi p_2^2)
\!-\!\xi {\cal{K}}^2 \right) \!
\hspace*{-5mm}
\nonumber \\
\left. \left.
+\xi (\xi-2) p_3^2 
\left(
2 (n-2) {\cal{K}} (p_1 p_2) (p_2 p_3)
-2 (n-1) p_1^2 p_3^2 (p_1 p_2) (p_2 p_3)
+{\cal{K}}^2 \right)
\right] 
\right\} , 
\hspace{8mm}
\end{eqnarray}
\begin{eqnarray}
\label{c(1,xi)}
c^{(1,\xi)}(p_3,p_2,p_1) =
-\frac{g^2 \; \eta}{(4\pi)^{n/2}} \; C_A \; \frac{1}{16 {\cal{K}}^2 p_1^2}
\hspace{82mm}
\nonumber \\
\times\! \left\{
\left[
\left( (n\!-\!3) \xi \!+\!2 \right) 
\left( \xi (n\!-\!1) (p_1 p_2)^2 (p_1 p_3)^2
-{\cal{K}} \left( 4 (p_1^2)^2+(p_1 p_3) 
( (\xi\!+\!2) p_1^2+3 \xi (p_1 p_2) ) \right)
\right)
\right. \right.
\nonumber \\
\left.
-\xi (\xi-2) (p_1 p_3) \left( (n-1) (p_1 p_2) (p_1 p_3)^2
                             +{\cal{K}} (2 p_1^2+3 (p_1 p_2)) \right) 
\right]
\; \left[ (p_2 p_3) \varphi+ \kappa_1 \right]
\nonumber \\
+\left[ \left( (n-3) \xi+2 \right) 
\left( \xi (p_1 p_3) (p_2 p_3) ((n-1) (p_1 p_2)^2+2 {\cal{K}})
                        +2 {\cal{K}} p_1^2 (p_2^2-(p_1 p_2)) \right)
\right.
\hspace{12mm}
\nonumber \\
\left.
+\xi (\xi\!-\!2) (p_1 p_3) 
\left( p_1^2 {\cal{K}}-(n\!-\!3) {\cal{K}} {\cal{Q}}
-(n\!-\!1) p_1^2 p_2^2 p_3^2 \right) 
\right]
 \left[ (p_1 p_3) \varphi+ \kappa_2 \right]
\nonumber \\
+ \left[ \left( (n-3) \xi+2 \right) 
\left( \xi (n-1) (p_1 p_2) (p_1 p_3)^2 (p_2 p_3)
\right. \right.
\hspace{64mm}
\nonumber \\
\left.
-{\cal{K}} \left( (\xi-2) p_1^2 (p_1^2-p_2^2)
       +\xi (p_2 p_3) (2 (p_1 p_3)-p_1^2) \right) \right)
\hspace{53mm}
\nonumber \\
\left.
+ \xi (\xi-2) (p_1 p_3) (p_2 p_3) 
\left( (n-3) {\cal{K}}-(n-1) p_1^2 p_3^2 \right)
\right]
\; \left[ (p_1 p_2) \varphi+ \kappa_3 \right]
\nonumber \\
+ 2 \varphi \; \left[ -\left( (n-3) \xi+2 \right) (p_1 p_2) 
\left( 2 p_1^2 (p_1^2-p_2^2) {\cal{K}}
       +\xi (n-1) (p_1 p_2) (p_1 p_3)^2 (p_2 p_3) \right)
\right.
\hspace{12mm}
\nonumber \\
\left. \left.
+\xi (p_1 p_3) (p_2 p_3)
\left( (\xi\!-\!2) (n\!-\!1) (p_1 p_2) (p_1 p_3)^2
       +\xi (n\!-\!2) {\cal{K}} (p_3^2-p_2^2) \right) 
\right]
\right\} ,
\hspace{12mm}
\end{eqnarray}
\begin{eqnarray}
\label{d(1,xi)}
d^{(1,\xi)}(p_3,p_2,p_1) =
-\frac{g^2 \; \eta}{(4\pi)^{n/2}} \; C_A \; \frac{1}{16 {\cal{K}}^2 p_3^2}
\hspace{82mm}
\nonumber \\
\times\! \left\{
\left[ \xi \left( (n-4) \xi+4 \right) (p_1 p_2) (p_1 p_3) 
           \left( (n-3) (p_1 p_3) (p_2 p_3)+p_3^2 (p_1 p_2) \right)
\right. \right.
\hspace{35mm}
\nonumber \\
- \left( (n-3) \xi+2 \right) p_3^2 
\left( \xi (p_2 p_3) (p_1^2 p_2^2-(n-2) (p_1 p_2) (p_1 p_3))
                                 +6 (p_1 p_2) {\cal{K}} \right)
\hspace{22mm}
\nonumber \\
\left.
+\xi (\xi-2) p_3^2 (p_2 p_3) (p_1^2 (p_2 p_3)-(n-2) (p_1 p_2)^2) \right]
\; \left[ (p_2 p_3) \varphi+\kappa_1 \right]
\nonumber \\
+ \left[ \xi ((n-4) \xi+4) (n-3) (p_1 p_2) (p_1 p_3) (p_2 p_3)^2
\right.
\hspace{67mm}
\nonumber \\
+\left( (n-3) \xi+2 \right) p_3^2 
\left( \xi (p_1 p_2) (p_2 p_3) ((n-3) (p_2 p_3)-2 p_2^2)
                                 -4 p_2^2 {\cal{K}} \right)
\hspace{29mm}
\nonumber \\
\left.
-(\xi\!-\!2) p_3^2 
\left( \xi (p_2 p_3) (p_1^2 p_2^2+(n\!-\!1) p_2^2 (p_1 p_2)+(p_1 p_2)^2)
                           +(\xi\!-\!2) p_2^2 {\cal{K}} \right)  \right]
\left[ (p_1 p_3) \varphi+\kappa_2 \right]
\nonumber \\
+ \left[ \xi \left( (n-4) \xi+4 \right) (n-4) (p_1 p_3) (p_2 p_3) {\cal{K}}
+\xi (\xi-2) p_3^2 (p_1 p_2) \left( (n-1) p_3^2 (p_2 p_3)-{\cal{K}} \right)
\right.
\nonumber \\
\left.
-\left( (n-3) \xi+2 \right) p_3^2 (p_2 p_3) 
\left( \xi (n-1) p_2^2 (p_1 p_3)+(\xi+6) {\cal{K}} \right)
\right]
\; \left[ (p_1 p_2) \varphi+\kappa_3 \right]
\nonumber \\
+ 2 \varphi \;
\left[-\left( (n-4) \xi+4 \right) (p_1 p_2) (p_2 p_3) {\cal{K}} 
\left( \xi (n-3) (p_1 p_3)+(\xi-2) p_3^2 \right)
\right.
\hspace{28mm}
\nonumber \\
+\left( (n-3) \xi+2 \right) p_3^2 
\left( \xi (p_1 p_2) (p_2 p_3) (2 {\cal{K}}+n p_2^2 (p_1 p_3))
                                 -2 (p_2^2)^2 (p_1 p_3)^2 \right)
\hspace{13mm}
\nonumber \\
\left. \left.
+(\xi-2) p_3^2 (p_1 p_2) (p_1 p_3) (p_2 p_3) 
\left( \xi (n-1) (p_1 p_2) +2 (\xi-1) p_2^2 \right) \right]
 \right\} ,
\hspace{17mm}
\end{eqnarray}
\begin{eqnarray}
\label{e(1,xi)}
e^{(1,\xi)}(p_3,p_2,p_1) =
-\frac{g^2 \; \eta}{(4\pi)^{n/2}} \; C_A \; \frac{1}{16 {\cal{K}}^2 p_1^2}
\hspace{82mm}
\nonumber \\
\times\! \left\{
\left[
\left( (n-3) \xi+2 \right) 
\left( -\xi (n-1) p_3^2 (p_1 p_2)^2 (p_1 p_3)
                        +{\cal{K}} p_1^2 (2 (p_1 p_3)-\xi p_3^2) \right)
\right. \right.
\hspace{30mm}
\nonumber \\
\left.
+\xi (\xi-2) (n-1) p_3^2 (p_1 p_2) (p_1 p_3)^2
         +\xi^2 (n-2) {\cal{K}} p_3^2 \left( 2 p_1^2+3 (p_1 p_2) \right)
\right]
\; \left[ (p_2 p_3) \varphi+\kappa_1 \right]
\nonumber \\
+ \left[ \left( (n-3) \xi+2 \right) (p_2 p_3) 
\left( -\xi (n-1) p_1^2 p_2^2 p_3^2
                             +{\cal{K}} (\xi (n-3) p_3^2+2 p_1^2) \right)
\right.
\hspace{30mm}
\nonumber \\
\left.
+\xi (\xi-2) p_3^2 
\left( (n-1) p_1^2 p_2^2 p_3^2+{\cal{K}} ((n-3) {\cal{Q}}-p_1^2) \right)
\right]
\left[ (p_1 p_3) \varphi+\kappa_2 \right]
\nonumber \\
+\left[ \left( (n-3) \xi+2 \right) 
\left( -\xi (n-1) p_1^2 p_2^2 p_3^2
       +{\cal{K}} ((\xi+2) p_1^2-(n-3) \xi {\cal{Q}}) \right)
\right.
\hspace{31mm}
\nonumber \\
\left.
+\xi (\xi-2) (p_2 p_3) \left( (n-1) (p_1 p_3)^2+2 {\cal{K}} \right) 
\right]
\; p_3^2 \; \left[ (p_1 p_2) \varphi+\kappa_3 \right]
\nonumber \\
+ 2 (p_2 p_3) \; \varphi \;
\left[ \left( (n-3) \xi+2 \right) (p_1 p_3) 
\left( \xi (n-1) p_3^2 (p_1 p_2)^2-2 {\cal{K}} p_1^2 \right)
\right.
\hspace{37mm}
\nonumber \\
\left.  \left.
-\xi (\xi-2) (n-1) p_3^2 (p_1 p_2) (p_1 p_3)^2
-\xi^2 (n-2) {\cal{K}} p_3^2 (p_3^2-p_2^2) 
\right]
\right\} .
\hspace{4mm}
\end{eqnarray}

From the expressions (\ref{a(1,xi)})--(\ref{e(1,xi)}), one can see
that again, as in the three-gluon functions 
(\ref{A(1,xi)})--(\ref{H(1,xi)}), 
the values $\xi=0$ and $\xi=-4/(n-4)$ are distinguished.
Putting $\xi=0$, we get rid of the momenta squared
in the denominators (only ${\cal{K}}$ may remain, 
cf. eqs.~(\ref{a(1,0)})--(\ref{e(1,0)}) ), 
while for $\xi=-4/(n-4)$ one of the momenta squared, $p_1^2$, still 
survives in the denominators
of the $c$ function, eq.~(\ref{c(1,xi)}), and the $e$ function,
eq.~(\ref{e(1,xi)}). This does not matter,
since (i) the $c$ and $e$ functions do not contribute to the
Ward--Slavnov--Taylor identity for the three-gluon vertex, eq.~(\ref{WST}),
and (ii) for the proper ghost-gluon vertex, eq.~(\ref{ghg}),
we get an extra $p_1^2$ in the numerator as a result of contracting
with ${p_1}^{\mu}$ (since the tensor structures corresponding to the
$c$ and $e$ functions contain ${p_1}_{\mu}$, see eq.~(\ref{BC-ghg}) ).
If one wants to put some of the momenta squared equal to zero in other
gauges, one should carefully consider the appropriate limit and expand the
functions in the numerator, in exactly the same way as it was described 
in Section~4 for the case of the three-gluon vertex.

In the limit $n\to 4$, the only divergent function in the one-loop 
ghost-gluon vertex is the $a$ function (\ref{a(1,xi)}), also since this
is the only function which is present at the zero-loop level, 
eq.~(\ref{ghg0}). The ultraviolet-divergent part of the $a$ function 
(\ref{a(1,xi)}) is 
\begin{equation}
\label{a_UV}
a^{(1,UV)} = \frac{g^2 \; \eta}{(4\pi)^{2-\varepsilon}} \;
C_A \; {\textstyle{1\over2}} \; (1-\xi) \; \kappa^{(UV)} ,
\end{equation}
where, as in eq.~(\ref{A_UV}), $\kappa^{(UV)} = 1/\varepsilon + \ldots$
corresponds to the divergent part of the function $\kappa$, 
eq.~(\ref{kappa}). Therefore, by analogy with (\ref{A_CT}), the
counterterm contribution is given by\footnote{See also the discussion
of $\mu_{DR}$, ${\overline{g}}^2$, etc.
in section~3.5.}
\begin{equation}
\label{a_CT}
a^{(1,CT)} = -\frac{{\overline{g}}^2}{(4\pi)^2} \;
\frac{C_A}{2} (1-\xi)
\left( \frac{1}{\varepsilon} + R \right)
= -\frac{g^2 \; \eta}{(4\pi)^{2-\varepsilon}} \;
\frac{C_A}{2} (1-\xi)
\left( \frac{1}{\varepsilon} + R \right) + {\cal{O}}(\varepsilon) ,
\end{equation}
where 
${\overline{g}}^2 \equiv g^2 e^{-\gamma \varepsilon} (4\pi)^{\varepsilon}$, and
$R$ is the renormalization-scheme constant ($R=0$ in the
$\overline{\mbox{MS}}$ scheme, which is achieved by a suitable extraction
of the over-all factor in eq.~(\ref{a_CT}) ).  
In particular, there is no singularity in the one-loop ghost-gluon vertex
in the Landau gauge ($\xi=1$), see e.g.\ ref.~\cite{MarPag}.

\appendix
\section*{Appendix E: Results for $p_3^2=0$ in arbitrary gauge}
\setcounter{equation}{0}
\renewcommand{\thesection}{E}

For arbitrary $\xi$, the results for the gluon and ghost loop contributions 
to the scalar functions (\ref{BC-ggg}) are\footnote{We also consider
non-equivalent permutations of the arguments.}
\begin{eqnarray}
A^{(1, \xi)}(p_1^2, p_2^2; 0)
= - \frac{g^2 \; \eta}{(4\pi)^{n/2}} \; C_A \;
\frac{1}{64 (n-1) (n-4)}
\hspace{65mm}
\nonumber \\
\times \! \left\{
(n-1) (n-4) \xi \left( 4+(n-4)\xi \right) \;
\left[ \frac{p_2^2}{p_1^2} \kappa_1 + \frac{p_1^2}{p_2^2} \kappa_2 \right]
\right.
\hspace{40mm}
\nonumber \\
+ (n\!-\!4) \left[ 16(3n\!-\!2) + 4\xi (n\!-\!1)(10n\!-\!33)
- 5\xi^2 (n\!-\!1)(n\!-\!4) \right] \;
\left[ \kappa_1 + \kappa_2 \right]
\nonumber \\
\left.
-16 (n-1)\left( n+(n-3)\xi \right) \; (\delta_{12})^{-1} \;
\left[ \kappa_1 - \kappa_2 \right] \right\} ,
\hspace{20mm}
\end{eqnarray}
\begin{eqnarray}
A^{(1, \xi)}(0, p_1^2; p_2^2)
= - \frac{g^2 \; \eta}{(4\pi)^{n/2}} \; C_A \;
\frac{1}{64 (n-1) (n-4) (n-6)}
\hspace{52mm}
\nonumber \\
\times \! \left\{
\frac{p_2^2}{p_1^2} \;
\kappa_1 (n-1) (n-4) (n-6) \; \xi \left( 4 + (n-4) \xi \right)
\right.
\hspace{56mm}
\nonumber \\
+ \kappa_1 (n-4) (n-6) \left[ 16 (3n-2) + 4\xi (n-1)(10n-33)
                     - 5 \xi^2 (n-1)(n-4) \right]
\hspace{5mm}
\nonumber \\
+ 2 \kappa_2 \; (n-1) (n-4) (n-6) \;
\xi \left( 4 (n-3) - (n-4) \xi \right)
\hspace{45mm}
\nonumber \\
- 8 (n-1)(n-4) \; \kappa_2
\left[ (\delta_{12})^{-1} (n-6) \left( 2 + (n-3)\xi \right)
       + (\delta_{12})^{-2} \; \xi \left( n + (n-3)\xi \right)
\right]
\nonumber \\
- 4  (n-1) \; \left[ \kappa_1 - \kappa_2 \right] \;
\left[ (\delta_{12})^{-1} \; (n-6)
    \left( 4(n\!+\!1)+ \xi (5n\!-\!18) -\xi^2 (n\!-\!3) \right)
\right.
\nonumber \\
\left. \left.
+ (\delta_{12})^{-2} \; \left( 4 + (n-4) \xi \right) \;
\left( (n-3)\xi - n+6 \right)
- 2 \; (\delta_{12})^{-3} \xi \left( n + (n-3)\xi \right)
\right] \right\} ,
\hspace{4mm}
\end{eqnarray}
\begin{eqnarray}
B^{(1, \xi)}(p_1^2, p_2^2; 0)
= - \frac{g^2 \; \eta}{(4\pi)^{n/2}} \; C_A \;
\frac{1}{64 (n-1) (n-4) (n-6)}
\hspace{52mm}
\nonumber \\
\times \! \left\{
(n-1) (n-4) (n-6) \;\xi \left( 4+(n-4)\xi \right) \;
\left[ \frac{p_2^2}{p_1^2} \kappa_1 - \frac{p_1^2}{p_2^2} \kappa_2 \right]
\right.
\hspace{26mm}
\nonumber \\
-8\; (n-1)(n-4) \;
\left[ \kappa_1 + \kappa_2 \right] \; (\delta_{12})^{-1} \;
\xi \left( n + (n-3)\xi \right)
\hspace{28mm}
\nonumber \\
+ \left[ \kappa_1 \!-\! \kappa_2 \right]
\left[ 16(n\!-\!6)(4n^2\!-\!21n\!+\!14)
+ 4 \xi (n\!-\!1)(n\!-\!6)(10n^2\!-\!77n\!+\!152)
\right.
\nonumber \\
\left. \left.
-\xi^2 (n-1)(n-6)(5n^2-48n+104)
+ 16 \; (\delta_{12})^{-2} \; (n-1) \xi \left( n + (n-3)\xi \right)
\right]
\right\} ,
\hspace{4mm}
\end{eqnarray}
\begin{eqnarray}
B^{(1, \xi)}(0, p_1^2; p_2^2)
= - \frac{g^2 \; \eta}{(4\pi)^{n/2}} \; C_A \;
\frac{1}{64 (n-1) (n-4) (n-6)}
\hspace{52mm}
\nonumber \\
\times \! \left\{
- (n-1) (n-4) (n-6) \;\xi \left( 4+(n-4)\xi \right) \;
\left[ \frac{p_2^2}{p_1^2} \kappa_1
     + 2 \frac{p_1^2}{p_2^2} \kappa_2 \right]
\right.
\hspace{20mm}
\nonumber \\
- 8 (n\!-\!1)(n\!-\!4)  \kappa_2
\left[ (n\!-\!6) \left( 2 +\! (n\!-\!3) \xi \right)
\left( 1 + (\delta_{12})^{-1} \right)
+ (\delta_{12})^{-2} \; \xi \left( n \!+\! (n\!-\!3) \xi \right)
\right]
\nonumber \\
- (n-4)(n-6) \; \kappa_1
\left[ 16(4n-3) + 20\xi (n-1)(2n-7) - 5\xi^2 (n-1)(n-4) \right]
\nonumber \\
+ 4(n-1) \; \left[ \kappa_1 - \kappa_2 \right] \;
\left[ (\delta_{12})^{-1} \; (n-6)
\left( 4 + 3\xi (n-2) + \xi^2 (n-3) \right)
\right.
\hspace{12mm}
\nonumber \\
\left. \left.
- (\delta_{12})^{-2} \; \left( 4+(n-4)\xi \right)
\left( (n-3)\xi -n+6 \right)
+ 2 (\delta_{12})^{-3} \; \xi \left( n + (n-3) \xi \right)
\right]  \right\} ,
\hspace{4mm}
\end{eqnarray}
\begin{eqnarray}
C^{(1, \xi)}(p_1^2, p_2^2; 0)
= - \frac{g^2 \; \eta}{(4\pi)^{n/2}} \; C_A \;
\frac{1}{32 (n-1) (n-4) (p_1^2-p_2^2)}
\hspace{49mm}
\nonumber \\
\times \! \left\{
\left[ \frac{p_2^2}{p_1^2} \kappa_1 - \frac{p_1^2}{p_2^2} \kappa_2 \right]
\; (n-1) (n-4) \; \xi \; \left( 4 + (n-4)\xi \right)
\right.
\hspace{47mm}
\nonumber \\
\left.
+ \left[ \kappa_1\! -\! \kappa_2 \right]
 \left[ 16 n(n\!-\!4) + 4\xi (n\!-\!1)(6n^2\!-\!41n\!+\!72)
 - 3\xi^2 (n\!-\!1)(n\!-\!4)^2 \right]
\right\} ,
\hspace{5mm}
\end{eqnarray}
\begin{eqnarray}
C^{(1, \xi)}(0, p_1^2; p_2^2)
= - \frac{g^2 \; \eta}{(4\pi)^{n/2}} \; C_A \;
\frac{1}{32 (n-1) (n-4) (n-6) \; p_1^2 \; (p_1^2-p_2^2)}
\hspace{30mm}
\nonumber \\
\times \!\left\{ \frac{}{}
8 (n-1) p_1^2 \left[ \kappa_1 - \kappa_2 \right]
\left[ -(n-6) \left( 6 + (n-5) \xi \right)
\right. \right.
\hspace{42mm}
\nonumber \\
\left.
+ (n-6) (\delta_{12})^{-1} \left( 6 + \xi (n-7) - \xi^2 (n-3) \right)
- 2 (\delta_{12})^{-2} \; \xi \; \left( n + (n-3)\xi \right)
\right]
\nonumber \\
+ 8 (n-1)(n-4) (\delta_{12})^{-1} \; (p_1^2 + p_2^2) \; \kappa_1 \;
\xi \left( n + (n-3) \xi \right)
\hspace{45mm}
\nonumber \\
+ (n-4)(n-6) \kappa_1
\left[ -\frac{(p_2^2)^2}{p_1^2} (n-1) \xi \left( 4 + (n-4) \xi \right)
+ 16 \left( n p_1^2 - (4n-3) p_2^2 \right)
\right.
\hspace{6mm}
\nonumber \\
\left. \left.
+ 2 \xi (n-1) \left( (12n-38) p_1^2 - 4 (5n-17) p_2^2 \right)
+ \xi^2 (n-1)(n-4) (-3 p_1^2 + 4 p_2^2)
\frac{}{} \right]
\right\} ,
\hspace{4mm}
\end{eqnarray}
\begin{eqnarray}
F^{(1, \xi)}(p_1^2, p_2^2, 0)
= - \frac{g^2 \; \eta}{(4\pi)^{n/2}} \; C_A \;
\frac{1}{16 (n-1) (n-4) (n-6) \; (p_1^2-p_2^2)}
\hspace{34mm}
\nonumber \\
\times \! \left\{
\left[ \frac{1}{p_1^2} \kappa_1 - \frac{1}{p_2^2} \kappa_2 \right]
 (n-1)(n-4) \xi \left( 4 + (n-4)\xi \right)
\left( 4\xi (n-3) - 5 (n-6) \right)
\right.
\nonumber \\
+ \frac{2}{p_1^2-p_2^2} \;
\left[ (n-4) \left[ \kappa_1 + \kappa_2 \right]
- 2 (\delta_{12})^{-1} \; \left[ \kappa_1 - \kappa_2 \right]
\right]
\hspace{10mm}
\nonumber \\
\times \left[
8 (n-6)(4n-7) + 4 \xi (n-1)(n-6)(5n-11)
\right.
\hspace{30mm}
\nonumber \\
\left. \left.
+ \xi^2 (n-1) (5n^2-60n+108) - 2 \xi^3 n (n-1)(n-3)
\right]
\right\} ,
\hspace{4mm}
\end{eqnarray}
\begin{eqnarray}
F^{(1, \xi)}(0, p_1^2; p_2^2)
= - \frac{g^2 \; \eta}{(4\pi)^{n/2}} \; C_A \;
\frac{1}{32 (n-1) (n-4) (n-6) \; p_1^2 \; (p_1^2-p_2^2)}
\hspace{31mm}
\nonumber \\
\times \! \left\{
-\frac{2}{p_1^2-p_2^2}
\left[ (n-6) \frac{(p_2^2)^2}{p_1^2} \kappa_1
       +4(n-3)\xi \frac{(p_1^2)^2}{p_2^2} \kappa_2 \right]
(n-1)(n-4) \xi \left( 4 + (n-4)\xi \right)
\right.
\nonumber \\
+ (n-4) \kappa_1
\left[ \left( {}^{} 32(n-6)(3n+1)+4\xi(n-1)(n-6)(20n-63)
\right. \right.
\hspace{25mm} \nonumber \\
\left.
         -3\xi^2 (n-1)(n-4)(n+2) - 4 \xi^3 (n-1)(n-3)(n-4)
\right)
\nonumber \\
- (\delta_{12})^{-1} \left( {}^{} 32(n-6)(n+2) + 4\xi (n-1)(7n+30)
\right.
\hspace{30mm} \nonumber \\
\left.
       +\xi^2 (n-1)(3n^2+50n-24) - 4\xi^3 (n-1)(n-3)(n-16)
\right)
\nonumber \\
\left.
- 24 (\delta_{12})^{-2}
\left( 8(n-6) + 8\xi (n-1) + \xi^2 (n-1)(3n-2) + 2\xi^3 (n-1)(n-3)
\right) \right]
\nonumber \\
+ 16  p_1^2 \; 
\frac{\kappa_1 - \kappa_2}{p_1^2 - p_2^2}
\times \left[ \left( 4(n-6)(n^2-12n+17) - 2\xi (n-1)(n-6)(3n-10)
\right. \right.
\nonumber \\
\left.
+ \xi^2 (n-1)(n-5)(n^2-6n+12) + 2\xi^3 (n-1)(n-3)(n-5) \right)
\nonumber \\
+ (\delta_{12})^{-1} \left( 16 (n-4)(n-6) + 4 \xi (n-1)(4n-15)
\right.
\hspace{40mm} \nonumber \\
\left.
+ \xi^2 (n-1)(n-4)(5n-3) + 3\xi^3 (n-1)(n-3)(n-4) \right)
\nonumber \\
\left. \left.
+ 3 (\delta_{12})^{-2}
\left( 8(n-6) + 8\xi (n-1) + \xi^2 (n-1)(3n-2) + 2\xi^3 (n-1)(n-3)
\right) \right] \right\} ,
\hspace{4mm}
\end{eqnarray}
\begin{eqnarray}
H^{(1, \xi)}(p_1^2, p_2^2, 0)
= - \frac{g^2 \; \eta}{(4\pi)^{n/2}} \; C_A \;
\frac{1}{16 (n-1) (n-4) (n-6) \; (p_1^2-p_2^2)}
\hspace{34mm}
\nonumber \\
\times \! \left\{
- \left[ \frac{p_2^2}{p_1^2} \kappa_1 - \frac{p_1^2}{p_2^2} \kappa_2 \right]
 (n-1)(n-4) \xi \left( 4 + (n\!-\!4)\xi \right)
\left( 2\xi (n\!-\!3) - 3 (n\!-\!6) \right)
\right.
\nonumber \\
+ 2 (\delta_{12})^{-1} \;
\left[ (n-4) \left[ \kappa_1 + \kappa_2 \right]
      - 2 (\delta_{12})^{-1} \; \left[ \kappa_1 - \kappa_2 \right]
\right]
\nonumber \\
\times \left[ 24(n-6) + 2\xi (n-1)(n+12) + \xi^2 (n-1)(11n-12)
              + 6 \xi^3 (n-1)(n-3) \right]
\nonumber \\
+ \left[ \kappa_1 - \kappa_2 \right]
\left[ 16(n-6)(n^2\!-\!2n\!-\!2) + 4\xi (n-1)(n-6)(6n^2\!-\!37n\!+\!60)
\right.
\nonumber \\
\left. \left.
- \xi^2 (n-1) (7n^3-88n^2+376n-552) - 8 \xi^3 (n-1)(n-3)(n-5)
\right] \right\} ,
\hspace{4mm}
\end{eqnarray}
where $\kappa_i \equiv \kappa(p_i^2)$, see eq.~(\ref{kappa}), while 
the coefficients $\eta$ and $C_A$ are defined by eqs.~(\ref{eta}) 
and (\ref{C_A}), respectively.

\appendix
\section*{Appendix F: Results for $p_1^2=p_2^2=0$ 
                                        in arbitrary gauge}
\setcounter{equation}{0}
\renewcommand{\thesection}{F}

The scalar functions $U_i$ corresponding to the decomposition
of the three-gluon vertex in this limit are defined by eq.~(\ref{ggg_U_i}).
Using the expressions for the one-loop contributions to the $A,B,C,F$ 
and $H$ functions, eqs.~(\ref{A(1,xi)(0,0;p^2)})--(\ref{H(1,q)(0,0;p^2)}),
and the representations (\ref{U_1})--(\ref{U_7}),           
we get the following results for the one-loop
contributions to the $U_i$ functions:
\begin{eqnarray}
U_1^{(1,\xi)}(p^2) =  - \frac{g^2 \; \eta}{(4\pi)^{n/2}} \; C_A \;
\frac{1}{8(n-1)(n-4)} \; \kappa
\hspace{65mm}
\nonumber \\
\times
\left\{ 4(2n^2-15n+19) + 2\xi (n-1)(n-3)(4n-17) -\xi^2 (n-1)(n-4)^2
\right\},
\end{eqnarray}
\begin{eqnarray}
U_2^{(1,\xi)}(p^2) = \frac{g^2 \; \eta}{(4\pi)^{n/2}} \; C_A \;
\frac{1}{4(n-1)(n-4)} \; \kappa
\hspace{65mm}
\nonumber \\
\times
\left\{ 2(3n^2-17n+8) + \xi (n-1)(n-4)(8n-27) -\xi^2 (n-1)(n-4)^2
\right\},
\end{eqnarray}
\begin{eqnarray}
U_3^{(1,\xi)}(p^2) = \frac{g^2 \; \eta}{(4\pi)^{n/2}} \; C_A \;
\frac{1}{16 (n-1)(n-4)} \; \kappa
\hspace{68mm}
\nonumber \\
\times
\left\{ 4(n-4)(n-5) + 2\xi (n-1)(2n^2-19n+36) -\xi^2 (n-1)(n^2-8n+20)
\right\},
\end{eqnarray}
\begin{eqnarray}
U_4^{(1,\xi)}(p^2) = \frac{g^2 \; \eta}{(4\pi)^{n/2}} \; C_A \;
\frac{1}{4 (n-1)(n-4)(n-6) p^2} \; \kappa
\hspace{50mm}
\nonumber \\
\times
\left\{ 6(n-2)(n-6)(n-7) + \xi (n-1)(n-6)(2n^2-23n+40)
\right.
\hspace{20mm}
\nonumber \\
\left.
+\xi^2 (n-1)(n-2)(n^2-15n+57)
-\xi^3 (n-1)(n-3)(n-8)
\right\},
\end{eqnarray}
\begin{equation}
U_5^{(1,\xi)}(p^2) = \frac{g^2 \; \eta}{(4\pi)^{n/2}} \; C_A \;
\frac{1}{4(n-1)(n-4) p^2} \; \kappa
\left\{ 4 (n-4)^2 + \xi (n-1)(n-6) \right\} ,
\end{equation}
\begin{eqnarray}
U_6^{(1,\xi)}(p^2) = - \frac{g^2 \; \eta}{(4\pi)^{n/2}} \; C_A \;
\frac{1}{16 (n-1)(n-4) p^2} \; \kappa
\hspace{50mm}
\nonumber \\
\times
\left\{ 16 n(n-4) + 4\xi (n-1)(6n^2-41n+72) - 3\xi^2 (n-1)(n-4)^2
\right\} ,
\end{eqnarray}
\begin{eqnarray}
U_7^{(1,\xi)}(p^2) = - \frac{g^2 \; \eta}{(4\pi)^{n/2}} \; C_A \;
\frac{1}{16 (n-1)(n-4) p^2} \; \kappa
\hspace{55mm}
\nonumber \\
\times
\left\{ 8(5n^2\!-\!25n\!+\!2) + 4\xi (n\!-\!1)(8n^2\!-\!59n\!+\!112)
- \xi^2 (n\!-\!1)(3n^2\!-\!24n\!+\!40)
\right\} ;
\end{eqnarray}
\begin{equation}
U_1^{(1,q)}(p^2) = \frac{g^2 \; \eta}{(4\pi)^{n/2}} \; 2 N_f T_R \;
\frac{n(n-3)}{(n-1)(n-2)} \; \kappa ,
\end{equation}
\begin{equation}
U_2^{(1,q)}(p^2) = - \frac{g^2 \; \eta}{(4\pi)^{n/2}} \; 4 N_f T_R \;
\frac{n-2}{n-1} \; \kappa ,
\end{equation}
\begin{equation}
U_3^{(1,q)}(p^2) = \frac{g^2 \; \eta}{(4\pi)^{n/2}} \; 4 N_f T_R \;
\frac{1}{(n-1)(n-2)} \; \kappa ,
\end{equation}
\begin{equation}
U_4^{(1,q)}(p^2) = \frac{g^2 \; \eta}{(4\pi)^{n/2}} \; 4 N_f T_R \;
\frac{n+2}{(n-1)(n-2) p^2} \; \kappa ,
\end{equation}
\begin{equation}
U_5^{(1,q)}(p^2) = - \frac{g^2 \; \eta}{(4\pi)^{n/2}} \; 4 N_f T_R \;
\frac{n-4}{(n-1)(n-2) p^2} \; \kappa ,
\end{equation}
\begin{equation}
U_6^{(1,q)}(p^2) = \frac{g^2 \; \eta}{(4\pi)^{n/2}} \; 4 N_f T_R \;
\frac{n-2}{(n-1) p^2} \; \kappa ,
\end{equation}
\begin{equation}
U_7^{(1,q)}(p^2) =  \frac{g^2 \; \eta}{(4\pi)^{n/2}} \; 4 N_f T_R \;
\frac{n^2-4n+6}{(n-1)(n-2) p^2} \; \kappa ,
\end{equation}
where, as usual, $\kappa\equiv \kappa(p^2)$.
Comparison with the definition of the functions $F_i$ given 
in eq.~(29) of \cite{BF} shows that 
the functions $U_1, U_2, U_3, U_4, U_5, U_6$ and $U_7$ are
proportional to $F_2, F_3, F_1, F_6, F_5, F_4$ and $F_7$,
respectively.

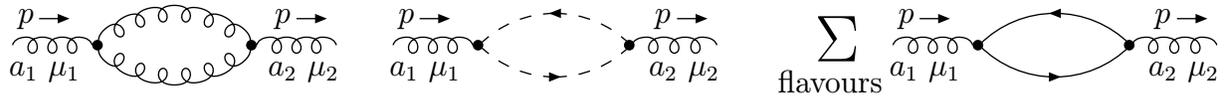
\end{document}